\documentclass[preprint]{revtex4}
\usepackage{amsmath,amssymb}
\usepackage{graphicx,natbib}
\usepackage{dcolumn}
\usepackage{bm}\usepackage{hyperref}
\usepackage{cleveref}

\begin{document}	
	\title{Li\'{e}nard Type Nonlinear Oscillators and Quantum Solvability}
	\author{V. Chithiika Ruby and M. Lakshmanan}
	\address{Department of Nonlinear Dynamics, School of Physics,
		Bharathidasan University, Tiruchirapalli - 620 024, India.}
		
\begin{abstract}
Li\'{e}nard-type nonlinear oscillators with linear and nonlinear damping terms exhibit diverse dynamical behavior in both the classical and quantum regimes. In this paper, we consider examples of various one-dimensional Li\'{e}nard type-I and type-II oscillators. The associated Euler-Lagrange equations are divided into groups based on the characteristics of the damping and forcing  terms. The Li\'{e}nard type-I oscillators often display localized solutions, isochronous and non-isochronous oscillations and are also precisely solvable in quantum mechanics in general, where the ordering parameters play an important role. These include Mathews-Lakshmanan and Higgs oscillators. However, the classical solutions of some of the nonlinear oscillators are expressed in terms of elliptic functions and have been found to be quasi-exactly solvable in the quantum region. The three-dimensional generalizations of these classical systems add more degrees of freedom, which show complex dynamics. Their quantum equivalents are also explored in this article. The isotonic generalizations of the non-isochronous nonlinear oscillators have also been solved both classically and quantum mechanically to advance the studies. The modified Emden equation categorized as Li\'{e}nard type-II exhibits isochronous oscillations at the classical level. This property makes it a valuable tool for studying the underlying nonlinear dynamics.  The study on the quantum counterpart of the system provides a deeper understanding of the behavior in the quantum realm as a typical PT-symmetric system.
\end{abstract}
%\pacs{03.65.-w, 03.65.Ge, 03.65.Fd}

\maketitle

\section{Introduction}
Li\'{e}nard type equations are second-order ordinary nonlinear differential equations that are primarily configured to describe   certain nonlinear electronic oscillatory circuits and later studied in the context of certain mechanical systems as well as other systems occurring in different branches of science and engineering \cite{lienard1928etude, lins2006lienard, jordan2007nonlinear}. These equations may be further  classified based on the presence of linear and nonlinear damping terms as suggested in Ref. \cite{lakshmanan2013generating} as follows
\begin{itemize}
\item Li\'{e}nard type-I or quadratic Li\'{e}nard type:
\begin{equation}
\ddot{x} + f(x) \dot{x}^2 + g(x) = 0,  \label{lie1}
\end{equation}
\item Li\'{e}nard type-II: 
\begin{equation}
\ddot{x} + f(x) \dot{x} + g(x) = 0,  \label{lie2}
\end{equation}
\item Li\'{e}nard type-III: 
\begin{equation}
\ddot{x} + f(x) \dot{x}^2 + g(x) \dot{x} + h(x) = 0, 
\end{equation}
\item Li\'{e}nard type-IV: 
\begin{equation}
\ddot{x} + f(x, t) \dot{x}^2 + g(x, t) \dot{x} + h(x, t) = 0, 
\end{equation}
\item Coupled versions of the above Li\'{e}nard type of dynamical systems: 
\begin{eqnarray}
\hspace{-0.5cm}\ddot{x} + f_1(x, y, t) \dot{x}^2 + f_2(x, y, t) \dot{y}^2 + f_3(x, y, t)\;\dot{x}\;\dot{y} + g_1(x, y, t)\;\dot{x} + g_2(x, y, t)\;\dot{y} + h_1(x, y, t) = 0,  \nonumber \\\\
\hspace{-0.5cm}\ddot{y} + f_4(x, y, t) \dot{x}^2 + f_5(x, y, t) \dot{y}^2 + f_6(x, y, t)\;\dot{x}\;\dot{y} + g_3(x, y, t)\;\dot{x} + g_4(x, y, t)\;\dot{y} + h_2(x, y, t) = 0, \nonumber\\
\end{eqnarray}
where  $f, g$ and $h$ are arbitrary functions of the variables indicated in the brackets and over dot denotes differentiation with respect to $t$. 
\end{itemize}

In the literature, many works have been devoted to studying the above types of nonlinear systems both from physical and mathematical perspectives. Certain nonlinear oscillators belonging to the above classes are of considerable physical interest, like the Mathews-Lakshmanan oscillator and Higgs oscillator, which belong to quadratic Li\'{e}nard type nonlinear oscillators. They describe the  dynamics of the harmonic oscillator in appropriate curved spaces  \cite{mathews1974unique, mathews1975quantum, higgs1979dynamical}. Similarly the motion of a particle on a rotating parabolic well and $k$-dependent polynomial belong to the class, Li\'{e}nard type-I \cite{venkatesan1997nonlinear, ruby2021classical}. A wide variety of nonlinear systems like the van der Pol oscillator \cite{van1920theory, van1934nonlinear}, Duffing oscillator \cite{kovacic2011duffing}, etc. come under the category of Li\'{e}nard type equation-II. Various mathematical techniques have been developed to study the integrability and  linearizability nature of the  above type of nonlinear systems. One finds that certain Li\'{e}nard type oscillators are solvable and also admit isochronous oscillations, whose frequencies are independent of their amplitudes \cite{mustafa2021isochronous}. Lie-group techniques, local and non-local transformations, etc., are employed to understand/ generate new isochronous systems \cite{pradeep2009dynamics, chandrasekar2012class, tiwari2015isochronous}. In another aspect, a systematic procedure has been developed to construct non-standard forms of Lagrangian/ Hamiltonian for specific nonlinear systems \cite{pradeep2009nonstandard,musielak2008standard}. Their corresponding quantized versions are also of considerable contemporary interest \cite{ruby2012exact}.  

In this study, we investigate various types of nonlinear dynamical systems that exhibit periodic solutions, both with and without amplitude dependency. We examine systems such as the Mathews-Lakshmanan oscillator and the Higgs oscillator, which have non-isochronous solutions, as well as systems with isochronous oscillations \cite{mathews1974unique, mathews1975quantum, higgs1979dynamical, mustafa2021isochronous}. We also explore systems that lead to elliptic functions in the classical regime, such as the $k$-dependent non-polynomial oscillator mentioned in \cite{ruby2021classical}. We investigate the dynamics of such  systems in the classical and quantum regimes. It is observed that the Hamiltonians corresponding to nonlinear systems governed by Li\'{e}nard type-I equations are in the form of position dependent mass. The position-dependent effective mass of physical systems has been the subject of recent research because of its many applications in studying the electronic properties of semiconductors \cite{bastard1992monographies, gonul2002exact, kocc2003systematic}, quantum dots, and quantum wells \cite{gora1969theory}.  While quantizing such Hamiltonians, the ordering parameters play an important role \cite{chithiika2015removal}. The quantum counterpart of the position dependent mass system possesses a linear energy spectrum if the classical solution is isochronous. Otherwise the energy spectrum is nonlinear that is quadratic in $n$ for exactly solvable quantum systems as observed in the cases of  Mathews-Lakshmanan oscillator and Higgs oscillator \cite{mathews1974unique, higgs1979dynamical}, and the same phenomenon is also observed for the general ordered form of the corresponding Hamiltonians, \cite{ruby2021classical, karthiga2017quantum}.  The study has been extended to consider three-dimensional generalizations of the oscillators, and the results show that the energy spectrum for non-isochronous oscillators continues to be quadratic in $n$.  The paper is organized as follows. In Sec. II, we provide a brief overview of the classical dynamics of the various Li\'{e}nard type I oscillators. We explore the quantum dynamics of the nonlinear oscillators in section III by examining exact and quasi-exact solvability through various mathematical techniques. The investigation also covers the effects of nonlinearity on the energy spectrum and eigenstates of the oscillators. In Sec. IV, we explore the classical and quantum dynamics of Li\'{e}nard type II nonlinear oscillators. Finally, in the conclusion section,  we briefly summarize  the classical and quantum studies on the nonlinear oscillators. 

\section{Classical dynamics of  Li\'{e}nard type I oscillators}
Very often, the nature of the solution of the classical dynamics of a system determines the solvability of its quantum counterpart, e.g., the Kepler particle dynamics and the hydrogen atom. Classical system that admits elementary periodic solutions is quantum mechanically completely solvable for certain choices of the ordering parameters, whereas the system admitting elliptic functions is generally found to be quasi-exactly solvable or not solvable analytically \cite{ruby2021classical}. 
In this section, we will discuss about the quadratic Li\'{e}nard type nonlinear oscillators possessing isochronous and non-isochronous solutions and also localized solutions. To do so,  we classify the quadratic Li\'{e}nard type equations based on the associated Lie point symmetry groups as (i) the maximal (eight parameter) symmetry group and  (ii) the non-maximal (three, two and one parameter) symmetry groups \cite{tiwari2013classification}. 

\subsection{Li\'{e}nard Type - I oscillators:  Isochronous oscillations}
 It is observed that the nonlinear systems of the form (\ref{lie1}),  admitting eight parameter Lie point symmetry elements are linearizable under coordinate transformations if the functions $g(x)$ and $f(x)$ of Eq. (\ref{lie1}) are related by the condition \cite{tiwari2015isochronous}
\begin{equation}
g(x) = e^{-\int f(x) dx} \left[g_1 \int e^{\int f(x) dx}\;\; dx + g_2 \right], \label{trans} 
\end{equation}
where $g_1$ and $g_2$ are arbitrary constants or equivalently by the differential equation $g' + fg = g_1$. Here $f(x)$ is any arbitrary function of $x$ that defines the system. For this choice (\ref{trans}), one can consider the transformation \cite{tiwari2013classification}, 
\begin{equation}
X(x) =\frac{g_1}{\omega^2_0}\left[\int\;e^{\int f(x) dx} dx + \frac{g_2}{g_1} \right], \label{trans1} 
\end{equation}
which transforms Eq. (\ref{lie1}) to be the equation of a linear harmonic oscillator with $g_1 = \omega^2_0$ as 
\begin{equation}
\ddot{X} + \omega^2_0 X = 0.  \label{ho}
\end{equation}
It results in the oscillatory solution, $X = A \sin(\omega_0 t + \delta)$, with the period ${\displaystyle T = \frac{2\pi}{\omega_0}}$ and $A, \omega_0$ and $\delta$ are arbitrary constants. These solutions are referred to as isoperiodic  or isochronous. We can consider several typical examples as follows. 

\subsubsection{\bf Exponential type -  $f(x) = \lambda$:}
The choice, $f(x) = \lambda = constant$ fixes $g(x)$ as $g(x) = g_1 + g_2 e^{-\lambda x}$. Choosing $g_1 = \frac{\omega_0^2}{\lambda}$ and $g_2 = -\frac{\omega_0^2}{\lambda}$, equation (\ref{lie1}) becomes 
\begin{eqnarray}
\ddot{x} + \lambda \dot{x}^2 + \frac{\omega^2_0}{\lambda} \left(1 - e^{-\lambda\;x}\right) = 0, 
\label{eq1}
\end{eqnarray}
whose solution is isochronous of the form
\begin{equation}
\hspace{-0.5cm} x = \frac{1}{\lambda} \ln\left(1 - \lambda A \sin(\omega_0 t + \delta)\right), \qquad 0 < A < \frac{1}{\lambda}, \label{sol1}
\end{equation}
where $A$ and $\delta$ are arbitrary constants. 
The corresponding Lagrangian is 
\begin{eqnarray}
L &=& \frac{\lambda^2}{2}\;e^{2\;\lambda\;x}\dot{x}^2  - \frac{\omega^2_0}{2}\left(1 - e^{\lambda\;x}\right)^2. \label{ham1}
\end{eqnarray}

Then the canonically conjugate momentum is 
\begin{eqnarray}
p = \lambda^2\;e^{2\;\lambda x}\; \dot{x}
\label{momentum1-iso}
\end{eqnarray}
so that the Hamiltonian becomes
\begin{eqnarray}
H = \frac{1}{2\lambda^2} e^{-2\;\lambda\;x} p^2  + \frac{\omega^2_0}{2}\left(1 - e^{\lambda\;x}\right)^2. 
\label{hamiltonian1-iso}
\end{eqnarray}

\begin{figure}
\vspace{1cm}
\begin{center}
\hspace{-1cm}\includegraphics[width=0.45\linewidth]{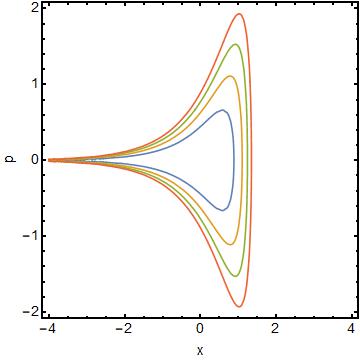}
\end{center}
\caption{The phase portrait $(x, p)$ of (\ref{hamiltonian1-iso}) for $\lambda = 1$ and $\omega^2_0 = 0.1.$ Different color trajectories represent different values of energy $H = E = 0.1, 0.2, 0.3$ and $0.4$.}
\label{exp}
\vspace{2cm}
\end{figure}

The dynamics of the system (\ref{hamiltonian1-iso}) is depicted as the $(x-p)$-phase portrait in Figure \ref{exp}.

\subsubsection{\bf A non-singular inverse form:}
Fixing $f(x) = -\frac{2\lambda}{1+\lambda x}$ yields the equation (\ref{lie1}) as 
\begin{eqnarray}
\ddot{x} - \frac{2\;\lambda}{1 + \lambda x} \dot{x}^2 +\omega^2_0 \left(x + \lambda x^2\right) = 0,  
\label{eq2}
\end{eqnarray}
and the corresponding Lagrangian takes the form, 
\begin{eqnarray}
L = \frac{\dot{x}^2}{2\;(1 + \lambda x)^4}  - \frac{\omega^2_0\; x^2}{(1 + \lambda x)^2} . \label{ham2}
\end{eqnarray}
The Hamiltonian can be obtained as 
\begin{eqnarray}
H = \frac{(1 + \lambda x)^4}{2} p^2 +  \frac{\omega^2_0\; x^2}{(1 + \lambda x)^2}, \label{hamiltonian2-iso}
\end{eqnarray}
where ${\displaystyle p = \frac{\dot{x}}{(1 + \lambda x)^4}}$ is the canonical conjugate momentum. 
The solution for Eq. (\ref{eq2}) is obtained as 
\begin{equation}
\hspace{-0.5cm} x = \frac{A \sin(\omega_0 t + \delta)}{1 - \lambda A \sin(\omega_0 t  + \delta)}, \qquad 0 < A < \frac{1}{\lambda}, \label{sol2}
\end{equation}
where $A$ and $\delta$ are arbitrary constants. 

\begin{figure}
\vspace{1cm}
\begin{center}
\hspace{-1cm}\includegraphics[width=0.45\linewidth]{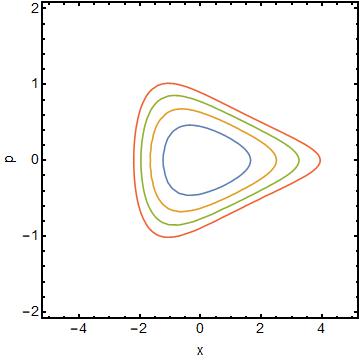}
\end{center}
\caption{The phase portrait of system (\ref{hamiltonian2-iso}) for $\lambda = 0.1$ and $\omega^2_0 = 0.1$ is shown at various energy eigenvalues $H = E = 0.1, 0.2, 0.3, 0.4$, represented by different colors.}
\label{h1-iso}

\end{figure}

The phase portrait of the system (\ref{hamiltonian2-iso}) is depicted in Figure \ref{h1-iso}.

Recently, Mustafa has observed that a class of nonlinear oscillators of the Li\'{e}nard type-I, including the systems (\ref{hamiltonian1-iso})  and (\ref{hamiltonian2-iso}), possesses isochronous oscillations  \cite{mustafa2021isochronous}. He proposed a methodology of obtaining isochronous nonlinear oscillators by considering the position dependent mass (PDM) equivalence of the Lagrangian as 
\begin{equation}
\hspace{-0.5cm} L = \frac{1}{2} m(x) \dot{x}^2 - Q(x)\omega^2_0 x^2. \label{geneal}
\end{equation}
He also observed that the negative of the gradient of the potential force field of such PDM oscillator is equal to the time derivative of the Noether momentum, $\pi(x) = \sqrt{m(x)} \dot{x}$.  

\subsubsection{\bf A non-singular inverse square form:}
Here $f(x)$ and $g(x)$ are replaced by the deformation function $Q(x)$ and mass function $m(x)$ by Mustafa \cite{mustafa2021isochronous}. For the choice of $Q(x) = \frac{1}{1 \pm \lambda^2 x^2}$, one can obtain the Lagrangian, 
\begin{eqnarray}
L &=& \frac{\dot{x}^2}{2\;(1 \pm \lambda^2 x^2)^3}  - \frac{\omega^2_0\; x^2}{(1 \pm \lambda^2 x)^2}, \label{ham3}
\end{eqnarray}
and the dynamical equation becomes 
\begin{eqnarray}
\ddot{x} \mp \frac{3\;\lambda^2\; x}{1 \pm \lambda^2 x^2} \dot{x}^2 +\omega^2_0\; x\;\left(1 \pm \lambda^2 x^2\right) = 0. \label{eq3}
\end{eqnarray}
Eq. (\ref{eq3}) admits the isochronous solution as 
\begin{equation}
\hspace{-0.5cm} x = \frac{A \sin(\omega_0 t + \delta)}{\sqrt{1 \mp \lambda^2 A^2 \sin^2(\omega_0 t  + \delta)}}, \label{sol3}
\end{equation}
where $A, \omega^2_0$ and $\delta$ are arbitrary constants. Here, for negative sign, the amplitude of the oscillations is restricted by $0 < A < \frac{1}{\lambda}.$

\begin{figure}
\vspace{1cm}
\begin{center}
\hspace{-1cm}\includegraphics[width=0.45\linewidth]{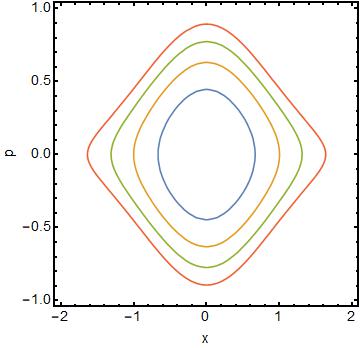}
\end{center}
\caption{The phase portrait of (\ref{hamiltonian2-cubic}) is shown for $\lambda = 0.5$ and $\omega_0^2 = 0.5$ at different energy levels: $H = E = 0.1, 0.2, 0.3, \text{and } 0.4$, each represented by a different color.}
\label{h2-cubic}
\end{figure}

The corresponding Hamiltonian can be expressed as 
\begin{eqnarray}
H = \frac{(1 \pm \lambda^2 x^2)^3\;p^2}{2} + \frac{\omega_0^2 \; x^2}{2\;(1 \pm \lambda^2 x^2)}, \qquad p = \frac{\dot{x}}{(1 \pm \lambda^2 x^2)^3}. \label{hamiltonian2-cubic}
\end{eqnarray}
The phase portrait of the system (\ref{hamiltonian2-cubic}) is depicted in Figure \ref{h2-cubic}.

\subsubsection{\bf A singular type of deformation function:}

For the choice, ${\displaystyle Q(x) = \frac{1}{1 - \lambda x}}$, the Lagrangian takes the form, 
\begin{eqnarray}
L &=& -\frac{(\lambda x - 2)^2\;\dot{x}^2}{8\;(\lambda x-1)^3}  - \frac{\omega^2_0\; x^2}{(1 - \lambda x)}, \label{ham4}
\end{eqnarray}
and the equation of motion is,  
\begin{eqnarray}
\ddot{x} - \frac{\lambda\;(\lambda x - 4)}{2(\lambda x - 1)(\lambda x - 2)} \dot{x}^2 + \frac{2 \omega^2_0\; (\lambda x -1)}{\left(\lambda x -2\right)}x = 0, \label{eq4}
\end{eqnarray}
which admits the isochronous solution as 
\begin{equation}
x = x_{\pm} = \frac{A}{2} \sin(\omega_0 t + \delta)\left[-\lambda A \sin(\omega_0 t + \delta) \pm \sqrt{\lambda^2 A^2 \sin^2(\omega_0 t  + \delta) +4 } \right], \qquad 0 < A < \frac{1}{\lambda}, \label{sol4}
\end{equation}
where $A, \omega^2_0$ and $\delta$ are arbitrary constants. 

The Hamiltonian can be obtained from (\ref{ham4}) as
\begin{eqnarray}          
 H =  -\frac{2\;(\lambda x - 1)^3}{(\lambda x - 2)^2} p^2 + \frac{\omega_0^2}{2\;(1 - \lambda x)}x^2, \qquad p = -\frac{(\lambda x - 2)^2\;\dot{x}}{4\;(\lambda x-1)^3}.  \label{hamiltonian2-negative}
\end{eqnarray}

\begin{figure}
\vspace{2cm}
\begin{center}
\hspace{-1cm}\includegraphics[width=0.45\linewidth]{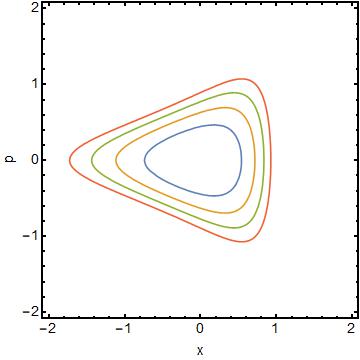}
\end{center}
\caption{The phase portrait of system (\ref{hamiltonian2-negative}) for $\lambda = 0.5, \omega^2_0 = 0.5$. Different color trajectories represent different values of energy $H = E = 0.1, 0.2, 0.3$ and $0.4$.}
\label{h2-negative}
\end{figure}

The dynamics of the system (\ref{hamiltonian2-negative}) is depicted in Figure \ref{h2-negative}.

\subsubsection{\bf A power-law type form:}
The deformation function, $Q(x) = a^2 x^{2 \nu}$, results in the Lagrangian of the form, 
\begin{eqnarray}
L &=&  a^2(\nu+1)^2\;x^{2\nu}\;\frac{\dot{x}^2}{2} - \frac{\omega^2_0\; a^2}{2}\; x^{2\nu+2},  \label{ham5}
\end{eqnarray}
which corresponds to the equation of motion,  
\begin{eqnarray}
\ddot{x} + \frac{\nu}{x} \dot{x}^2 + \frac{1}{\nu + 1} \omega^2_0 x = 0. \label{eq5}
\end{eqnarray}
One can obtain the corresponding isochronous solution as 
\begin{equation}
x = \left[\frac{A}{a}\sin(\omega_0 t + \delta)\right]^{1/(\nu+1)}, \qquad \nu \neq -1,  \label{sol4}
\end{equation}
where $A, \omega^2_0$ and $\delta$ are arbitrary constants. 

The corresponding Hamiltonian can be expressed as 
\begin{eqnarray}
H = \frac{x^{-2 \nu}\;p^2}{2\;a^2\;(\nu+1)^2} + \frac{\omega_0^2\; a^2\; x^2}{2}\; x^{2\nu + 2}, \qquad p = a^2(\nu+1)^2\;x^{2\nu}\;{\dot{x}}.\label{hamiltonian2-power}
\end{eqnarray}

\begin{figure}
\vspace{2cm}
\begin{center}
\hspace{-1cm}\includegraphics[width=0.45\linewidth]{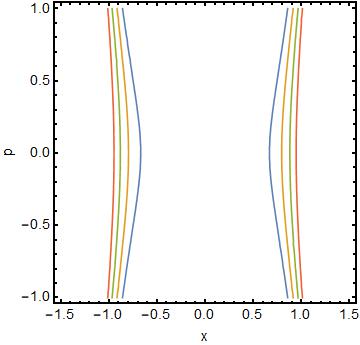}
\end{center}
\caption{The phase portrait of system (\ref{hamiltonian2-power}) for $\nu = a= \omega_0 = 1$. Various color trajectories correspond to distinct energy values, specifically $H = E = 0.1, 0.2, 0.3$, and $0.4$.}
\label{h2-power}
\end{figure}

The phase portrait of the system (\ref{hamiltonian2-power}) is depicted in Figure \ref{h2-power}.

The functions, $f(x)$ and $g(x)$, characterizing the nonlinear systems (\ref{eq1}), (\ref{eq2}), (\ref{eq3}), (\ref{eq4}) and (\ref{eq5}) satisfy the isochronous condition,    
\begin{equation}
g_x + f g = \omega^2_0 = constant, \label{iso-condition}
\end{equation}
where the suffix $x$ represents differentiation with respect to $x$, as pointed out in Ref. \cite{tiwari2013classification}. 

\subsection{Li\'{e}nard Type - I: Non-isochronous oscillations:}
Typical nonlinear systems characterized by the arbitrary functions $f(x)$ and $g(x)$  of Li\'{e}nard type equations (\ref{lie1}) do not satisfy the isochronous condition (\ref{iso-condition}). However, certain of these dynamical equations can be transformed into the form of the dynamical equation for a simple harmonic oscillator like system (or even anharmonic oscillator systems) by using appropriate transformations. These systems admit periodic oscillations with the amplitude dependent frequencies and so admit non-isochronous oscillations. It can also be shown that such systems possess minimum point symmetry elements. For example, Mathews-Lakshmanan oscillator possesses one point symmetry, namely the translation symmetry, \cite{tiwari2013classification}. Higgs oscillator and $k$-dependent non-polynomial oscillator also possess non-isochronous oscillations and single point symmetries, \cite{ruby2021classical}.

Classically, the kinetic energy term of the Mathews-Lakshmanan oscillator (MLO) depicts the dynamics of  a nonlinear $SU(2) X SU(2)$- chirally invariant Lagrangian expressed in Gasiorowicz-Geffen co-ordinates 
\cite{delbourgo1969infinities}, 
\begin{equation}
{\cal L} = \frac{1}{2}\left[(\partial_{\mu} {\bf \Phi}).(\partial_{\mu} {\bf \Phi})  + \frac{\lambda ( {\bf \Phi}.\partial_{\mu} {\bf \Phi})( {\bf \Phi}. \partial_{\mu} {\bf \Phi})}{(1 - \lambda {\bf \Phi}^2)}\right], 
\label{gasiro}
\end{equation}
where $\lambda = constant$, and is expressed in flat space $\bf{\Phi} \rightarrow {\bf q}$, 
\begin{equation}
L = \frac{1}{2}\left[\dot{\bf q}^2 +  \frac{\lambda ({\bf q}. \dot{\bf q})^2}{(1 -\lambda {\bf q}^2)}\right].  
\label{gasiroq}
\end{equation}
Higgs geometrically  interpretated the above nonlinear system as the one which  arises on the gnomonic projection onto the tangent plane from the center of the sphere \cite{higgs1979dynamical}.

In another coordinate system, namely Schwinger coordinates, the same nonlinear chiral system (\ref{gasiro}) is expressed as  
\begin{equation}
{\cal L} = \frac{1}{2}\left[(\partial_{\mu} {\bf \Phi}).(\partial_{\mu}  {\bf \Phi}) - \frac{k}{(1 + k {\bf \Phi}^2)}\left( {\bf \Phi}^2 (\partial_{\mu}{\bf \Phi}). (\partial_{\mu} {\bf \Phi}) +\frac{ ({\bf \Phi}. \partial_{\mu} {\bf \Phi})({\bf \Phi}. \partial_{\mu}  {\bf \Phi})}{(1 + k {\bf \Phi}^2)}\right)\right],   
\label{schwinger}
\end{equation}
where $k$ is constant \cite{delbourgo1969infinities}. 

The nonlinear systems associated with the Lagrangian (\ref{schwinger}), in the flat space, is related to the Higgs oscillator. In flat space, the Lagrangian takes the form 
\begin{equation}
L = \frac{1}{2}\left[\dot{\bf q}^2- \frac{k}{(1 + k {\bf q}^2)}\left( {\bf q}^2 {\bf {\dot q}}^2 +\frac{ ({\bf q}. \dot{\bf q})^2}{(1 + k{\bf q}^2)}\right)\right].  
\label{schwingerq}
\end{equation}

Similarly another possible choice of parametrization of the Lagrangian density (\ref{gasiro}) is expressed as 
\begin{eqnarray}
{\cal L} = \frac{1}{2}\left[(\partial_{\mu} {\bf \Phi}).(\partial_{\mu}  {\bf \Phi})\left(\frac{1}{(1 + \lambda^2\;{\bf \Phi}^2)^2} - 1\right)\right], \label{wein}
\end{eqnarray}
which is known as Lagrangian density expressed in terms of Weinberg coordinates \cite{delbourgo1969infinities} and it can be expressed in flat space as 
\begin{eqnarray}
L = -\frac{\lambda^2\;{\bf q}^2}{2\;(1+\lambda^2 {\bf q}^2)^2}\;\dot{\bf q}^2. \label{wein}
\end{eqnarray}

Cari\~{n}ena et al. \cite{carinena2004non,carinena2007quantum} provided a physical interpretation of the Mathews-Lakshmanan oscillator and the Higgs oscillator. They deduced these systems while analyzing the behavior of the harmonic oscillator on two-dimensional Riemann spaces ($M^2_k$) with constant curvature (that is, sphere $S^2$ and hyperbolic plane ${\cal H}^2$). In this unified approach, the curvature $k$ was treated as a parameter and the corresponding Lagrangian of the harmonic oscillator potential in terms of  geodesic polar coordinates $(R, \Phi)$ on $M^2_k$ which is expressed as \cite{carinena2004non,carinena2007quantum}
\begin{eqnarray}
{\cal L}(k) = \frac{1}{2}\left(v^2_R + S^2_k(R) v^2_{\Phi}\right) - \frac{1}{2}\omega^2_0 T^2_k(R), \label{carl}
\end{eqnarray}
where ${\displaystyle T_k = \frac{S_k}{C_k}}$, $S_k$ and $C_k$ are $k$-dependent trigonometric functions. They are defined as 
\begin{eqnarray}
C_k &=& \left \{
  \begin{tabular}{cc}
  $\cos(\sqrt{k}\;x),$ & $k>0$  \\
  1, & $k = 0$ \\
  $\cosh(\sqrt{-k}\;x),$ & $k<0$ 
  \end{tabular}
\right.\\
S_k &=& \left \{
  \begin{tabular}{cc}
  $\frac{1}{\sqrt{k}}\sin(\sqrt{k}\;x),$ & $k>0$  \\
  $x$, & $k = 0$ \\
  $\frac{1}{\sqrt{-k}}\sinh(\sqrt{-k}\;x),$ & $k<0$ 
  \end{tabular}
\right.
\end{eqnarray}
Here $k>0$ denotes the spherical plane, whereas  $k < 0$ denotes hyperbolic plane and $k = 0$ denotes the Euclidean plane. 

(1) On the transformation $(R,\Phi) \rightarrow (r', \phi)$ as $r' = T_k(R)$, $\Phi = \phi$, the Lagrangian ${\cal L}(k)$ becomes
\begin{eqnarray}
{\cal L}_{H}(k) = \frac{1}{2}\left(\frac{v^2_{r'}}{(1 + k r'^2)^2} + \frac{v^2_{\phi}}{(1 + k r'^2)}\right)- \frac{1}{2}\omega^2_0 r'^2, 
\label{higgs2}
\end{eqnarray}
which is known as the two dimensional Higgs oscillator \cite{higgs1979dynamical}. 

(2) Similarly, on the transformation $(R,\Phi) \rightarrow (r, \phi)$ as $r = S_k(R)$, $\Phi = \phi$ with $k = - \lambda$, the Lagrangian ${\cal L}(k)$ becomes
\begin{eqnarray}
{\cal L}_{M}(\lambda) = \frac{1}{2}\left(\frac{v^2_{r}}{(1 + \lambda r^2)^2} + r^2 v^2_{\phi}\right)-
\frac{\omega^2_0}{2} \frac{r^2}{(1 + \lambda r^2)}, 
\label{higgs2}
\end{eqnarray}
which is known as the two dimensional Mathews-Lakshmanan oscillator \cite{mathews1974unique}.

In this subsection, we discuss briefly about the classical dynamics of such non-isochronous nonlinear oscillators. 

\subsubsection{\bf Mathews-Lakshmanan oscillator:}
Mathews-Lakshmanan oscillator is characterized by a non-standard form of Lagrangian \cite{mathews1974unique}, 
\begin{equation}
L = \frac{1}{2}\frac{(\dot{x}^2 - \omega^2_0 x^2)}{(1 + \lambda x^2)}, \label{ml}
\end{equation}
where $\omega_0$ and $\lambda$ are parameters. It corresponds to the dynamical equation, 
\begin{eqnarray}
\ddot{x} - \frac{\lambda x }{1 + \lambda x^2} \dot{x}^2 + \frac{\omega^2_0 x}{1 + \lambda x^2} = 0, \qquad \lambda > 0,  \label{eq-ml}
\end{eqnarray}
and admits simple harmonic oscillatory solution \cite{mathews1974unique, mathews1975quantum}, 
\begin{eqnarray}
x(t) = A \sin(\Omega t  + \delta), \label{sol-ml}
\end{eqnarray}
with ${\displaystyle \Omega = \frac{\omega_0}{\sqrt{1 + \lambda A^2}}}$, for ${\displaystyle \frac{\omega^2_0}{2\lambda} > E}$, where $E$ is the energy of the system. 

The system (\ref{ml}) can be solved for both positive and negative values of $\lambda$.  Note that when $\lambda < 0$, the amplitude of the oscillations $ \;\displaystyle{|A |< |\lambda|^{-1/2}}.$ The solution (\ref{sol-ml}) is valid  for (\ref{eq-ml}) with the frequency ${\displaystyle \Omega = \frac{\omega_0}{\sqrt{1 - |\lambda| A^2}}}$. 

\begin{figure}
\vspace{1cm}
\begin{center}
\hspace{-1cm}\includegraphics[width=0.45\linewidth]{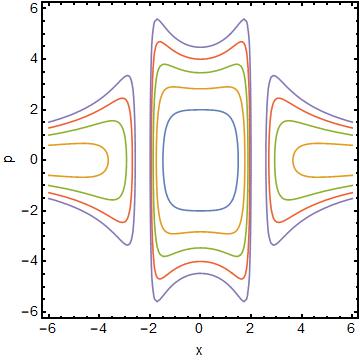}
\end{center}
\caption{The phase trajectories of the Hamiltonian system (\ref{ham-mlo}) with $\omega^2_0 = 1$, $\lambda = -0.2$ for various values of $E = H = 2, 4, 6, 8$ and $10$.}
\label{mlo-fig}
\end{figure}

The Hamiltonian corresponding to the equation of motion (\ref{eq-ml}) can be obtained as 
\begin{equation}
H = \frac{(1 + \lambda x^2)}{2} p^2 +  \frac{\omega^2_0 x^2}{(1 + \lambda x^2)}. \label{ham-mlo}
\end{equation}
The phase trajectories of the Hamiltonian (\ref{ham-mlo}) with $\omega^2_0 = \lambda = 0.1$ for various values of $E = H = 0.1, 0.2, 0.3$ and $0.4$ are plotted in Figure \ref{mlo-fig}.

The total energy of the system can be obtained as 
\begin{eqnarray} 
\hspace{2.5cm}E = H = \frac{1}{2}\;\frac{\omega_0^2 A^2}{(1 + \lambda A^2)}. \label{energy-mlo}
\end{eqnarray}

The nonlinear system (\ref{ml}) and its generalizations have been continuously studied for many aspects such as $3$-dimensional \cite{lakshmanan1975quantum} and $d$-dimensional generalizations \cite{ranada2002harmonic, lakshmanan2013generating} and rational extensions of the potentials  \cite{quesne2015generalized, quesne2018deformed}.
 
\subsubsection{\bf Higgs oscillator:}
The one-dimensional version of the Higgs oscillator is described by the Lagrangian, 
\begin{equation}
L = \frac{\dot{x}^2}{2(1 + k x^2)^2} - \frac{\omega^2_0}{2} x^2, 
\label{higg1d}
\end{equation} 
where $k$ is a parameter related to the curvature. The Higgs oscillator has also attracted considerable attention since its introduction to the literature \cite{carinena2004non}. 

The Euler-Lagrange's equation of motion for the system (\ref{higg1d}) is also of the form of the quadratic Li\'{e}nard type nonlinear equation as 
\begin{equation}
\ddot{x} - \frac{2 k x }{(1 + k x^2)} \dot{x}^2 + \omega^2_0\;{(1 + k x^2)^2}\;x = 0. 
\label{eom1}
\end{equation}
On implementation of the appropriate transformation and integration, one can get the solution, 
\begin{equation}
x(t) = \frac{A \sin(\Omega t + C)}{\sqrt{1 - k A^2 \sin^2(\Omega t + C)}},  
\label{xt}
\end{equation}
where ${\displaystyle |A|< \frac{1}{\sqrt{k}}}$ for $k>0$ and   $x(t)$ is periodic for all the values of $A$ when $k<0$ with  
\begin{eqnarray}
\hspace{2.7cm}E = \frac{\omega^2_0\;A^2}{1 - k A^2}, \qquad \Omega = \frac{\omega_0}{\sqrt{1 - k A^2}}.
\label{enhigg}
\end{eqnarray}
The Hamiltonian of the Higgs oscillator (\ref{higg1d}) can be shown to be of the form, 
\begin{equation}
H = \frac{(1 + k x^2)^2}{2}\;p^2 + \frac{\omega_0^2}{2} x^2. 
\label{ham-higgs}
\end{equation} 

\begin{figure}
\vspace{1cm}
\begin{center}
\hspace{-1cm}\includegraphics[width=0.45\linewidth]{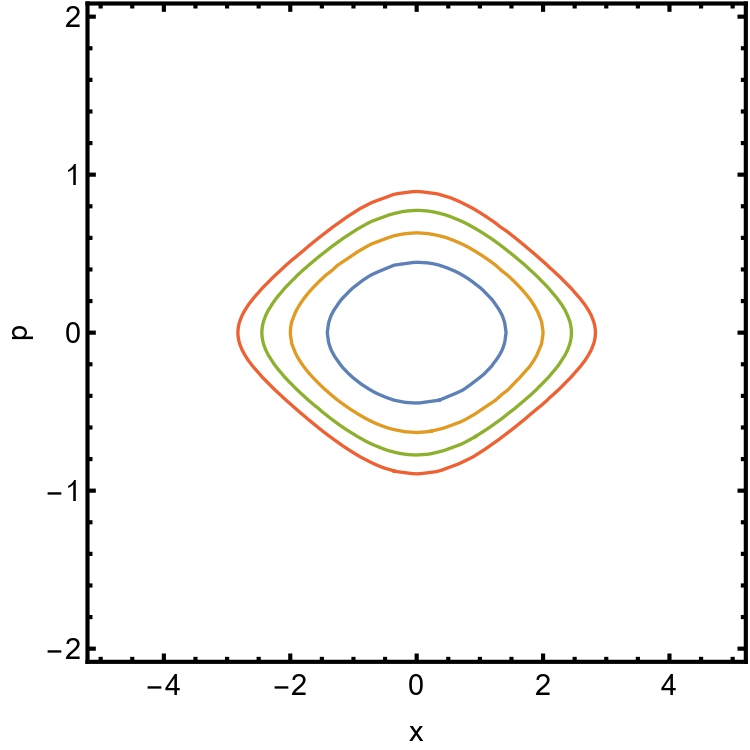}
\end{center}
\caption{The phase trajectories of the Hamiltonian system (\ref{ham-higgs}) with $\omega = k = 0.1$ for various values of $E = H = 0.1, 0.2, 0.3$ and $0.4$.}
\label{higgs-fig}
\end{figure}

The phase trajectories of the Hamiltonian (\ref{ham-higgs}) with $\omega^2_0 = \lambda = 0.1$ for various values of $E = H = 0.1, 0.2, 0.3$ and $0.4$ are plotted in Figure \ref{higgs-fig}.

The two dimensional Higgs oscillator and superintegrable generalizations of Higgs oscillator have been studied  using the algebraic technique that incorporates their superintegrability property \cite{bonatsos1994deformed, ballesteros2006universal, ballesteros2009maximal, ballesteros2013anisotropic}.

The fact that the explicit expression (\ref{xt}) of both the one dimensional Mathews-Lakshmanan and Higgs oscillator can be expressed in terms of simple trigonometric functions classically motivates one to study the exact solvability of their quantum versions, see below.

\subsubsection{\bf $k$-dependent non-polynomial oscillator:}
Recently, the one-dimensional version of the nonlinear chiral system (\ref{schwingerq}) corresponding to  a non-polynomial potential has been studied both classically and quantum mechanically \cite{ruby2021classical}. The  Lagrangian has been considered to be of the form
\begin{equation}
L = \frac{\dot{x}^2-\omega^2_0\;x^2}{2(1 + k x^2)^2}, 
\label{nlo1d}
\end{equation} 
where $k$ and $\omega_0$ are system parameters. The corresponding equation of motion is 
\begin{equation}
\ddot{x} - \frac{2 k x }{(1 + k x^2)} \dot{x}^2 + \omega^2_0 \frac{(1 - k x^2)\;x}{(1 + k x^2)} = 0, 
\label{eom2}
\end{equation}   
which is also of the form of the quadratic Li\'{e}nard type nonlinear equation, Eq. (\ref{lie1}).

The solution is obtained in terms of the Jacobian elliptic function as 
\begin{equation}
x(t) = A\;sn\left(\frac{\omega_0}{1 +  k A^2}\;t, m\right), 
\label{xtf}
\end{equation}
where the square of the modulus is 
\begin{eqnarray}
\hspace{2.9cm} m = k A^2.  
\end{eqnarray}

\begin{figure}
\vspace{1cm}
\begin{center}
\hspace{-1cm}\includegraphics[width=0.45\linewidth]{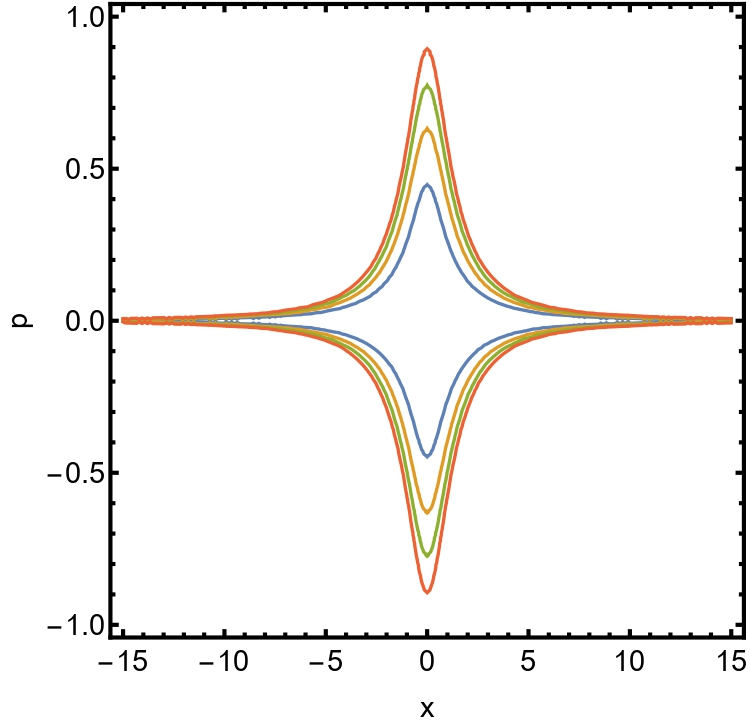}
\end{center}
\caption{The phase trajectories of the Hamiltonian system  with $\omega_0 = 0.1, k = 0.5$ for the values of $E = H = 0.1, 0.2, 0.3$ and $0.4$.}
\label{non-fig}
\end{figure}

The Hamiltonian of the system (\ref{nlo1d}) is of the form 
\begin{equation}
H = \frac{(1 + k x^2)^2}{2} p^2+\frac{\omega^2_0\;x^2}{2(1 + k x^2)^2}, 
\label{nlo1d}
\end{equation} 
whose phase-trajectories are displayed in Figure \ref{non-fig}.

The total energy of the system can be obtained as 
\begin{eqnarray}
\hspace{2.3cm}E = H = \frac{\omega^2_0\;A^2}{(1 + k A^2)^2}. 
\end{eqnarray}

As expected, since the classical solution (\ref{xtf}) is the Jacobian elliptic function $sn$, one finds that the quantum counterpart of the system becomes quasi exactly solvable as pointed out later in section III-B.

\subsection{\bf Li\'{e}nard type-I equation: Localized solution}
In this sub-section, we consider a quadratic Li\'enard type nonlinear oscillator which shows a special behavior at its classical level, that is, the system exhibits temporally localized solutions and possesses two Lie point symmetries \cite{tiwari2013classification}. It reads as 
\begin{equation}
\ddot{x} - \frac{2}{x}\;\dot{x}^2 + \lambda x^5 =0. 
\label{eqm}
\end{equation}
The corresponding Lagrangian is 
\begin{equation}
L = \frac{\dot{x}^2}{x^4} - \lambda x^2, 
\label{delta}
\end{equation}
and the Hamiltonian is 
\begin{equation}
H = \frac{x^4\;p^2}{4} + \lambda x^2, 
\label{hamiltonian}
\end{equation}
where the canonically conjugate momentum is 
\begin{eqnarray}
p = \frac{2 \dot{x}}{x^4}. 
\label{nlo-momentum}
\end{eqnarray}

It is of the position-dependent mass form, ${\displaystyle H = \frac{p^2}{2\;m(x)}+V(x)},$ where the mass profile is of the form 
\begin{equation}
m(x) = \frac{2}{x^4} \qquad \mbox{and} \quad V(x) = \lambda x^2. 
\label{mass}
\end{equation}
Here the mass is singular at $x = 0.$

Eq. (\ref{eqm}) admits the general solution,
\begin{equation}
x(t) = \frac{1}{\sqrt{\frac{\lambda}{C_1} + (C_2 + \sqrt{C_1}\;t)^2}}, 
\label{sol}
\end{equation}
where $C_2$ is the second integration constant. For $\lambda > 0$, we have a temporally localized solution. And for $\lambda < 0$, we  have  a singular solution when $t = \frac{1}{\sqrt{C_1}}\left(\sqrt{\frac{|\lambda|}{C_1}}-C_2\right)$ in which case we consider that  $C_1$ and $C_2$ are positive. 

\begin{figure}
\vspace{1cm}
\begin{center}
\hspace{-1cm}\includegraphics[width=0.45\linewidth]{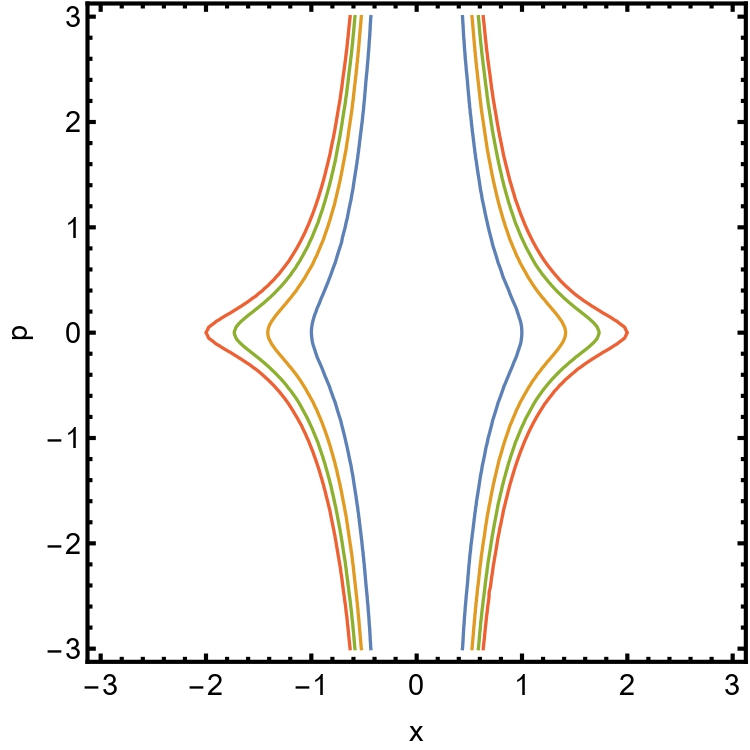}
\end{center}
\caption{The phase trajectories of the Hamiltonian system (\ref{nlo1d}) with $\lambda = 0.1$ for various values of $E = H = 0.1, 0.2, 0.3$ and $0.4$.}
\label{delta-fig}
\end{figure}

The phase-portrait of the system (\ref{hamiltonian}) is shown in the Figure \ref{delta-fig}.

The mass profile has a shape that resembles a $\delta$-function. A related model, which has been used to describe electron systems in $\delta$-doped semiconductors in the Thomas-Fermi field, has been shown to be quantum mechanically exactly solvable \cite{schulze2013position}.

\subsection{Isotonic generalization of non-isochronous nonlinear oscillators:}
Recently, Ra\~{n}ada studied the properties of Mathews-Lakshmanan oscillator by including the isotonic term, ${\displaystyle \frac{g}{2\;x^2}},$ which depicts the combined effect of two nonlinear terms on the classical and quantum dynamics of the isotonic Mathews-Lakshmanan oscillator \cite{ranada2014quantum}. With this motivation, we extend the studies to other non-isochronous nonlinear oscillators (\ref{higg1d}), (\ref{nlo1d}) and $\delta$-type mass systems (\ref{delta}) as well. 

\subsubsection{\bf Isotonic generalization of Mathews-Lakshmanan oscillator:}

The Lagrangian corresponding to the one-dimensional Mathews-Lakshmanan oscillator with an additional potential, ${\displaystyle \frac{g}{2\;x^2}}$, is considered in the following form, 
\begin{eqnarray}
L = \frac{\dot{x}^2}{2(1-\lambda x^2)}-\frac{\omega^2_0\; x^2}{2(1-\lambda x^2)}- \frac{g}{2\;x^2}, \qquad g > 0, \label{ml-iso}
\end{eqnarray}
where $\omega_0, \;\lambda$ and $g$ are arbitrary parameters, \cite{ranada2014quantum}. 

The equation of motion corresponding to the Lagrangian (\ref{ml-iso}) is, 
\begin{eqnarray}
\ddot{x} +\frac{\lambda\;x}{(1-\lambda x^2)} \dot{x}^2 + \frac{\omega^2_0\; x}{(1-\lambda x^2)} - g \left(\frac{1-\lambda x^2}{x^3}\right) = 0, \label{ml-iso-eq}
\end{eqnarray}
which has been solved for both positive and negative values of $\lambda$. Note that the sign in (\ref{ml-iso-eq}) corresponds to the case $\lambda = -|\lambda|$ in Eq. (\ref{ml}). 

{\bf (i) Positive values of $\lambda$}

For positive value of $\lambda$, $|x| < 1/{\sqrt{\lambda}}$, 
\begin{eqnarray}
x(t) = \left(\frac{1}{\Omega A}\right)\sqrt{(\Omega^2 A^4 - g)\;\sin^2(\Omega t  + \delta) + g}, \label{sol1-ml-iso} 
\end{eqnarray}
with the frequency, 
\begin{equation}
\Omega = \frac{\omega^2_0}{(1 - \lambda A^2)} + \frac{\lambda g}{A^2}.  \label{freq1-ml-iso}
\end{equation}
The frequency of oscillation is related with amplitude $A$.

{\bf (ii) Negative values of $\lambda$}

The system (\ref{ml-iso-eq}) admits bounded motion for $\lambda < 0$. It is obtained as 
\begin{eqnarray}
x(t) = \left(\frac{1}{\Omega A}\right)\sqrt{(\Omega^2 A^4 - g)\;\sin^2(\Omega t  + \delta) + g}, \label{sol1-ml-iso} 
\end{eqnarray}
with the frequency, 
\begin{equation}
\Omega^2 = \omega^2_0 - 2 |\lambda| E, \label{freq2-ml-iso}
\end{equation}
where the energy must be restricted as $E < E_b$ and $E_b = \frac{\omega^2_0}{2 |\lambda|}$.

Eq. (\ref{ml-iso-eq}) also admits unbounded motion as
\begin{eqnarray}
x(t) = \left(\frac{1}{\Omega A}\right)\sqrt{(\Omega^2 A^4 + g)\;\sinh^2(\Omega t  + \delta) + g}, \label{sol1-ml-iso} 
\end{eqnarray}
with the parameter, 
\begin{equation}
\Omega = \frac{\omega^2_0}{(|\lambda| A^2 - 1)} - \frac{|\lambda| g}{A^2}.  \label{freq1-ml-iso}
\end{equation}

\subsubsection{\bf Isotonic generalization of Higgs oscillator:}
Consider the Lagrangian corresponding to the one-dimensional Higgs oscillator in the presence of isotonic term, ${\displaystyle \frac{g}{2\;x^2}}$, 
\begin{eqnarray}
L = \frac{\dot{x}^2}{2(1+k x^2)^2}-\frac{\omega^2_0\; x^2}{2}- \frac{g}{2\;x^2}, \qquad g > 0, \label{higg1d-iso}
\end{eqnarray}
where, $\omega_0, k$ and $g$ are arbitrary parameters. Then the equation of motion can be written as 
\begin{eqnarray}
\ddot{x} - \frac{2 k\;x}{(1+k x^2)} \dot{x}^2 + \omega^2_0 x (1 + k x^2)^2 - g \left(\frac{(1 + k x^2)^2}{x^3}\right) = 0, \label{ml-eq}
\end{eqnarray}
which results in the solution
\begin{equation}
x(t) = \sqrt{\frac{(\Omega^2 A^4 - 2\;g) \sin^2(\Omega t + \delta) + g}{(\Omega^2 A^2-k g)-k(\Omega^2 A^4 - 2\;g)\;\sin^2(\Omega t + \delta)}}, \qquad k > 0, \label{sol-iso-higg1d}
\end{equation}
with frequency and energy, 
\begin{eqnarray}
\hspace{-0.5cm} \qquad \Omega^2 = \frac{\omega_0^2}{1-k A^2} - \frac{g k^2}{1- k A^2},  \quad 
E = \frac{\omega_0^2\;A^2}{2 (1-k A^2)}-\frac{g k}{2\;(1-k A^2)}(2-k A^2). 
\end{eqnarray}

\subsubsection{\bf Isotonic generalization of $k$-dependent non-polynomial oscillator:}

It is interesting to consider the isotonic generalization to the classical dynamics of the newly derived potential (\ref{nlo1d}). To do so, we consider the Lagrangian  as,
\begin{eqnarray}
L = \frac{\dot{x}^2}{2(1+k x^2)^2}-\frac{\omega^2_0\; x^2}{2\;(1+k x^2)^2}- \frac{g}{2\;x^2}, \qquad g > 0, \label{nlo1d-iso}
\end{eqnarray}
where, $k, \omega_0$ and $g$ are arbitrary parameters. The equation of motion associated with the Lagrangian (\ref{nlo1d-iso}) takes the form as
\begin{eqnarray}
\frac{\ddot{x}}{(1+k x^2)^2} - \frac{2 k\;x}{(1+k x^2)^3} \dot{x}^2 + \frac{\omega^2_0 x\;(1 - k x^2)}{(1 + k x^2)^3}- \frac{g}{x^3} = 0.  \label{nlo2-iso-eq}
\end{eqnarray}
By using the multiplication factor, ${\displaystyle \dot{x}}$, 	Eq. (\ref{nlo2-iso-eq}) can be integrated once to obtain   
\begin{eqnarray}
\frac{\dot{x}^2}{(1+k x^2)^2}+ \frac{\omega^2_0 x}{(1 + k x^2)^2}+\frac{g}{x^2} = constant = \epsilon. \label{nlo2d-iso-sol} 
\end{eqnarray}
The equation will lead to elliptic function solution on transformations. 

\subsubsection{\bf Isotonic generalization of $\delta$-type position dependent mass system:}
The inclusion of the isotonic term to the $\delta$-type potential  profoundly influences its classical dynamics as the system possess singularity at origin.  To study the property, we consider the Lagrangian, 
\begin{equation}
\hspace{-0.5cm} L = \frac{\dot{x}^2}{x^4} - \lambda x^2 - \frac{g}{2\;x^2}, \qquad g > 0,  \label{delta-iso}
\end{equation}
where $\lambda$ and $g$ are system parameters and are arbitrary.  The corresponding Euler-Lagrange's equation of motion is 
\begin{eqnarray}
\ddot{x} -\frac{2}{x} \dot{x}^2 + \lambda x^5 -\frac{g}{2} x= 0. \label{delta-iso-eq}
\end{eqnarray}
On solving Eq. (\ref{delta-iso-eq}), one can get 
\begin{eqnarray}
x(t) = \frac{\sqrt{g}}{\left[\sqrt{A^2+2 \lambda g} + A \cos(2\Omega t+\phi)\right]^{1/2}}, \label{delta-iso-sol}
\end{eqnarray}
where $\Omega = \sqrt{g/2}$ and $\epsilon = \sqrt{A^2 + 2 \lambda g}.$

The Hamiltonian from (\ref{delta-iso}) can be written as 
\begin{equation}
H = \frac{{x}^4}{4}\;p^2 + \lambda x^2 + \frac{g}{2\;x^2}, \qquad g > 0,  \label{ham-delta-iso}
\end{equation}

\begin{figure}
\vspace{1cm}
\begin{center}
\hspace{-1cm}\includegraphics[width=0.45\linewidth]{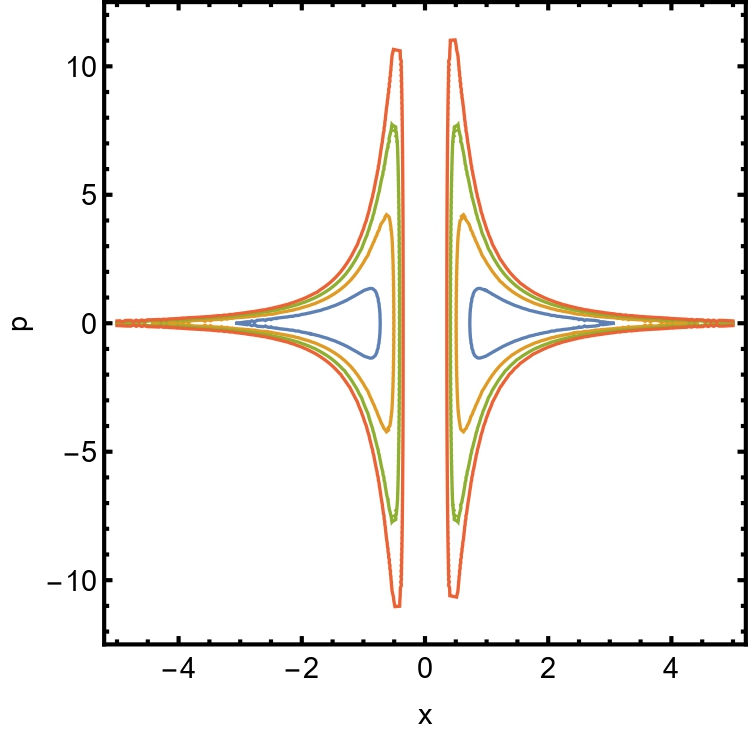}
\end{center}
\caption{The phase trajectories of the Hamiltonian system  (\ref{ham-delta-iso}) with $ g = 1$ and $\lambda= 0.1$ for various values of $E = H = 1, 2, 3$ and $4$.}\label{delta-iso-fig}
\end{figure}

The dynamics of the system (\ref{ham-delta-iso}) is depicted in Figure \ref{delta-iso-fig}.

\subsection{\label{3d} Three dimensional generalization of the nonlinear oscillators}

We have previously studied the behavior of nonlinear oscillators in one dimension or with one degree of freedom. Now, we examine their three-dimensional counterparts to gain a better understanding of their complex dynamics that are relevant to real-world circumstances. Specifically, we consider the three-dimensional generalization of non-isochronous nonlinear oscillators like Mathews-Lakshmanan, Higgs, and k-dependent non-polynomial oscillators, both at the classical and quantum levels. 

\subsubsection{Three dimensional Mathews-Lakshmanan oscillator}
The three-dimensional generalized version of the Mathews-Lakshmanan oscillator can be derived from the Lagrangian expressed in flat space (\ref{gasiroq}) by considering the generalized coordinates, ${\bf q}$ as $q_i,\;\;i = 1, 2, 3$, parameterizing a projective three-sphere of radius $\frac{1}{\sqrt{\lambda}}$. It takes the form for $\lambda = -\lambda$, 
\begin{equation}
\hspace{3.5cm}L = \frac{1}{2}\left[\dot{{\bf q}}^2 - \frac{\lambda ({\bf q}. \dot{{\bf q}})^2}{(1 + \lambda {{\bf q}}^2)} - \frac{\omega_0^2 ({\bf q})^2}{(1 + \lambda {\bf q}^2)}\right]. \nonumber \hspace{4.1cm} (\ref{gasiroq})
\end{equation}
It is $SU(2) \times SU(2)$ chirally invariant \cite{lakshmanan1975quantum}. The corresponding equation of motion is 
\begin{eqnarray}
\hspace{-0.5cm} \ddot{q}_i+\left[-\frac{\lambda\;({\bf q}.\ddot{\bf q})}{(1 + \lambda {\bf q}^2)} - \lambda \frac{(\dot{\bf q})^2}{(1 + \lambda {\bf q}^2)}+\frac{\lambda^2 ({\bf q}.\dot{\bf q})^2}{(1 + \lambda {\bf q}^2)^2} + \frac{\omega^2_0}{(1 + \lambda {\bf q}^2)^2}\right] q_i = 0, \qquad i = 1, 2, 3. 
\label{eom-mlo-3d}
\end{eqnarray}
Eq. (\ref{eom-mlo-3d}) has been solved for $\lambda = - |\lambda|$ and it admits the periodic solution as shown in Ref.\cite{lakshmanan1975quantum} by Lakshmanan and Eswaran. Now, expressing ${\bf q} = (r\sin{\theta}\cos{\phi}, r \sin{\theta}\sin{\phi}, r\cos{\theta})$, one can separate the radial part (that is for $r$) and angular parts (for ($\theta,\;\phi$)) of (\ref{eom-mlo-3d}) as 
\begin{eqnarray}
\frac{\ddot{r}}{(1 + \lambda r^2)} - \frac{\lambda r}{(1 + \lambda r^2)^2} \dot{r}^2 + \frac{\omega^2_0 r}{(1 + \lambda r^2)^2} - {r}
\left(\dot{\theta}^2 + \sin^2(\theta) \dot{\phi}^2\right) &=& 0,\label{rmlo}\\
\hspace{7cm} \dot{\theta}^2 +  \sin^2{\theta}\dot{\phi}^2 &=& \frac{C^2_2}{r^4}, \label{0mlo}\\
\hspace{7cm} r^2\;\sin^2{\theta}\;\dot{\phi} &=& C_1, \label{pmlo}
\end{eqnarray}
where $C_1$ and $C_2$ are constants. Using (\ref{0mlo}) and (\ref{pmlo}) in (\ref{rmlo}), one can obtain the solution for the radial part of (\ref{eom-mlo-3d}) as 
\begin{eqnarray}
r(t) &=& A \left[1 - \beta \sin^2(\Omega t + \zeta) \right]^{1/2}, \label{sol-mlo-3d}
\end{eqnarray}
where the parameters,
\begin{eqnarray}
\beta &=& 1 + \frac{1}{\lambda A^2}\left(1 + \lambda A^2 - \frac{\omega_0-\lambda^2 C^2_2}{\Omega^2}\right), \\
\Omega &=& \frac{\omega_0}{1 + \lambda A^2}-\frac{\lambda C^2_2}{A^2}, 
\end{eqnarray}
with 
\begin{eqnarray}
A &=& c + \sqrt{\left(c^2 + \frac{C^2_2}{\lambda C_3}\right)} , \\
c &=& -\frac{1}{2\lambda}\left(1 + \frac{\omega_0-\lambda^2 C^2_2}{C_3}\right), 
\end{eqnarray}
where $C_3$ is an integration constant. The oscillation (\ref{sol-mlo-3d}) is periodic for all the values of $\lambda$ for $\lambda >0$.  
For $\lambda < 0$, the periodic oscillations are restricted within the range, $0 < A < \frac{1}{\sqrt{|\lambda|}}$. 

%The quantum dynamics of the three dimensional generalization  have been understood well in the literature for particular orderings
\subsubsection{Three dimensional Higgs oscillator}
The Lagrangian corresponding to the Higgs oscillator (see (\ref{schwingerq})) is 
\begin{equation}
{L} = \frac{1}{2}\left[\frac{\dot{\bf q}^2}{(1 + k {\bf q}^2)}-\frac{ k\;({\bf q}. \dot{\bf q})^2}{(1 + k  {\bf q}^2)^2}- \omega^2_0 {\bf q}^2\right],  
\label{schwingerq-higg}
\end{equation}
where $\omega_0$ is the potential parameter, which results in the equation of motion, 
\begin{eqnarray}
\hspace{-0.5cm} \frac{\ddot{q}_i}{(1 + k {\bf q}^2)^2} - 2 k \frac{({\bf q}.\dot{\bf q})}{(1 + k {\bf q}^2)^3}\dot{q}_i + \left[\frac{2 k^2 ({\bf q}.\dot{\bf q})^2}{(1 + k {\bf q}^2)^3} + \omega^2_0\right] q_i = 0, \qquad i = 1, 2, 3. 
\label{eom-h3d}
\end{eqnarray}
We express the equation of motion (\ref{eom-h3d}) in polar coordinates $(r, \theta, \phi)$, as  
$q_1 = r \sin{\theta}\cos{\phi}, \; q_2  = r \sin{\theta} \sin{\phi}, \; q_3 = r \cos{\theta}$. 
The radial part (that is for $r$) and angular parts (for ($\theta,\;\phi$)) become, 
\begin{eqnarray}
\hspace{-0.5cm} \frac{\ddot{r}}{(1 + k r^2)^2} - \frac{2 k r}{(1 + k r^2)^3} \dot{r}^2 + \omega^2_0 r - \frac{r}{(1 + k r^2)^2}
\left(\dot{\theta}^2 + \sin^2(\theta) \dot{\phi}^2\right) &=& 0,\label{rhigg}\\
\hspace{-0.5cm} \hspace{8cm}\dot{\theta}^2 + \sin{\theta}^2 \dot{\phi}^2 &=& C^2_2\frac{(1 + k r^2)^2}{r^4}, \label{0higg}\\
\hspace{-0.5cm}\hspace{8cm}\frac{r^2\;\sin^2{\theta}\;\dot{\phi}}{(1 + k r^2)} &=& C_1, \label{phigg}
\end{eqnarray}
where $C_1$ and $C_2$ are constants. Equations (\ref{rhigg}-\ref{phigg}) admit the solution,  
\begin{equation}
\hspace{-0.5cm} \qquad r(t) =  A \left[\frac{\eta + \sin(\Omega t + \kappa) }{1 - k \eta A^2 - k A^2 \sin(\Omega t + \kappa)}\right]^{1/2}, 
\label{sol-r}
\end{equation}
where $\kappa$ is a constant and 
\begin{eqnarray}
\Omega &=& 2 \sqrt{\left(k^2\;C^2_2 + k C_3 + \omega_0^2\right)},  \\
A^2 &=& \Lambda = \frac{\sqrt{C^2_3 - 4 \omega^2_0 C^2_2}}{2 \left(k^2\;C^2_2 + k C_3 + \omega_0^2\right)}, \\
\eta &=& -\frac{\tau}{\Lambda} = \frac{C_3 + 2 k C_2^2}{\sqrt{C^2_3 -  4 \omega^2_0 C^2_2}}, 
\end{eqnarray}
where $C_3$ is an integration constant. When $k>0$, the solution (\ref{sol-r}) is periodic for $C^2_3 - 4 \omega^2_0 C^2_2>0$ and $C_3 > 0$. When $k<0$, the periodic solutions exist for $C^2_3 - 4 \omega^2_0 C^2_2>0$ and $\Omega>0$.

\subsubsection{Three dimensional generalization of $k$-dependent nonlinear oscillator}

The Lagrangian corresponding to the nonlinear oscillator (via (\ref{schwingerq})) takes the form,  
\begin{equation}
{L} = \frac{1}{2}\left[\frac{\dot{\bf q}^2}{(1 + k {\bf q}^2)} - \frac{k ({\bf q}. \dot{\bf q})^2}{(1 + k  {\bf q}^2)^2}- \frac{\omega^2_0 {\bf q}^2}{(1 + k \;{\bf q}^2)^2}\right],  
\label{schwingerq-nlor}
\end{equation}
where $\omega_0$ is the potential parameter, which results in the equation of motion, 
\begin{eqnarray}
\hspace{-0.5cm} \frac{\ddot{q}_i}{(1 + k {\bf q}^2)^2} - 2 k \frac{({\bf q}.\dot{\bf q})}{(1 + k {\bf q}^2)^2}\dot{q}_i + \left[\frac{2 k^2 ({\bf q}.\dot{\bf q})^2}{(1 + k {\bf q}^2)^3} -\frac{ k\;({\bf q}. \dot{\bf q})}{(1 + k  {\bf q}^2)^2}+ \frac{\omega^2_0 (1 - k  {\bf q}^2)}{(1 + k  {\bf q}^2)^3}\right] q_i = 0. 
\label{eom-nlo}
\end{eqnarray}
The equation of motion (\ref{eom-nlo}) in polar coordinates $(r, \theta, \phi)$  can be expressed as 
\begin{eqnarray}
\ddot{r} - \frac{2 k r}{(1 + k r^2)} \dot{r}^2 + \omega^2_0 r \left( \frac{1 - k r^2}{1 + k r^2}\right)-r
\left(\dot{\theta}^2 + \sin^2{\theta} \dot{\phi}^2\right) &=& 0,\label{rnlo2}\\
\hspace{5cm}\ddot{\theta} + \frac{2 \dot{r} \dot{\theta}}{r\;(1 + k r^2)} - \sin{\theta}\cos{\theta} \dot{\phi}^2 &=& 0, \label{0nlo2}\\
\hspace{8cm}\frac{r^2\;\sin^2{\theta}\dot{\phi}}{(1 + k r^2)} &=& C_1, \label{pnlo2}
\end{eqnarray}
where $C_1$ is constant. On substituting (\ref{pnlo2}) in (\ref{0nlo2}) and (\ref{rnlo2}), we can get 
\begin{eqnarray}
\frac{r^4 \dot{\theta}^2}{2\;(1 + k r^2)^2} + \frac{C^2_1}{2 \sin^2{\theta}} = \frac{C^2_2}{2}, \label{0enlo}\\
\frac{\ddot{r}}{(1 + k r^2)^2} - \frac{2 k r \dot{r}^2}{(1 + k r^2)^3} + \omega^2_0 r \left( \frac{1 - k r^2}{1 + k r^2}\right)- \frac{C^2_2}{r^3} = 0, \label{renlo}
\end{eqnarray}
where $C_2$ is a constant. Integrating twice, we can obtain the solution in terms of elliptic functions, though complicated to express explicitly. So we do not write the form here.

\section{Quantum Solvability of the Li\'{e}nard type-I systems}

In this section, we discuss about the quantum solvability of the various quadratic Li\'{e}nard type systems analyzed for their classical dynamics earlier. To do so, we quantize the associated Hamiltonians, which are in principle of non-standard forms, that is not simply kinetic energy plus potential energy. The Hamiltonians associated with the quadratic Li\'{e}nard type nonlinear oscillator equation (\ref{lie1}) are observed as position dependent mass ones ($m(x)$) which means that the mass of the corresponding system is dependent on position.  Hence, the mass function no longer commutes with the momentum in the quantum regime, and  ordering ambiguity arises while quantizing the position dependent mass Hamiltonian. Consequently, one may introduce the most general ordered form of the Hamiltonian  
operator as \cite{trabelsi2013classification}
\begin{equation}
\hat{H} = \frac{1}{2}\sum_{i = 1}^N w_i m^{\alpha_i} \hat{p} m^{\beta_i} \hat{p} m^{\gamma_i} + V(x),  \quad \hat{p} = -i\;\hbar\frac{d}{d x}. 
\label{geo1d}
\end{equation}
The form (\ref{geo1d}) allows many number of possible mixtures of the fundamental term, $m^{\alpha} \hat{p} m^{\beta} \hat{p} m^{\gamma}$, where $\hat{p}$ is the momentum operator. Here, the term $N$ represents any positive integer, and the ordering parameters $(\alpha_i, \beta_i, \gamma_i)$ are required to satisfy the constraint $\alpha_i + \beta_i + \gamma_i = -1,\;i = 1, 2, 3, \ldots, N$. Further, in (110) $w_i$'s are real and represent weight functions, summing up to $1$. This form provides a complete classification of Hermitian and non-Hermitian orderings, as stated in \cite{trabelsi2013classification}.

The kinetic energy operator of $\hat{H}$ in equation \eqref{geo1d} generally exhibits non-Hermitian characteristics and encompasses $2N$ independent ordering parameters. In ref. \cite{chithiika2015removal}, it is demonstrated that the inclusion of the mass function provides a pathway to establish a connection with its Hermitian counterpart via a similarity transformation:
\begin{equation}
\hat{H}_{\text{her}} = m^{\eta} \hat{H}_{\text{non}} m^{-\eta}.  \label{hermitian} 
\end{equation}
Here, the quantity $2\eta = \bar{\gamma} - \bar{\alpha}$, where $\bar{X}=(\bar{\alpha}, \bar{\beta}, \bar{\gamma})$ represents the weighted mean value defined as $\bar{X} = \sum_{i = 1}^N w_i X_i$, and $-\bar{\beta} = 1 + \bar{\alpha} + \bar{\gamma}$ (Ref. \cite{trabelsi2013classification}). By establishing a correspondence between the non-Hermitian Hamiltonian and its Hermitian counterpart,  through a similarity transformation, we effectively map the dynamics of the non-Hermitian system onto a Hermitian framework. Hence, the non-Hermitian ordered Hamiltonian possesses a real energy spectrum \cite{chithiika2015removal}. Consequently,  this property is known as quasi-Hermiticity, which implies that the system exhibits Hermitian-like behavior despite its non-Hermitian nature.

The  Hermitian Hamiltonian  (\ref{hermitian}) for a potential $V(x)$ can be expressed as 
\begin{eqnarray}
\hspace{-0.5cm}\hat{H}_{her} &= &\frac{1}{2}\hat{p}\frac{1}{m}\hat{p} + \frac{\hbar^2}{2}\left[\left(\frac{\bar{\alpha}+\bar{\gamma}}{2}\right) \frac{d^2}{d x^2}\left(\frac{1}{m}\right) + \left(\overline{\alpha\gamma}+ \frac{1}{4} (\bar{\gamma} - \bar{\alpha})^2\right)  \left(\frac{d}{dx}\left(\frac{1}{m}\right)\right)^2 m\right]\nonumber \\ 
\hspace{-0.5cm} & & \hspace{11cm}+ V(x).  
\label{geham_heo1d}
\end{eqnarray}

We utilize the coordinate representation ${\displaystyle \hat{p} = - i \hbar \frac{d}{d x}}$ in equation \eqref{geham_heo1d}. Consequently, we obtain the time-independent Schr\"{o}dinger equation for the Hermitian ordered Hamiltonian \eqref{geham_heo1d} in the following form:  
\begin{eqnarray}
\hspace{-0.5cm}\psi{''} - \frac{m{'}}{m} \psi{'} + 
\left(\left(\frac{\bar{\alpha}+\bar{\gamma}}{2}\right)  \frac{m{''}}{m}-\left(\overline{\alpha\gamma}+\bar{\gamma}+\bar{\alpha}+\frac{1}{4}(\bar{\gamma} - \bar{\alpha})^2\right)  \frac{m{'^2}}{m^2}\right) \psi \nonumber \\ 
\hspace{0.5cm}+ \frac{2 m}{\hbar^2}\left(E -V(x)\right)\psi  = 0, 
\label{seg_he}
\end{eqnarray}
where ${\displaystyle ' = \frac{d}{dx}}$ and we seek for the solution for arbitrary values of the ordering parameters ${\bar \alpha}$ and ${\bar \beta}$ \cite{chithiika2015removal}.

In contrast to the conventional way of expressing the Hamiltonian (\ref{geo1d}), researchers have recently used Dunkl formalism \cite{dunkl1991integral} in the quantum framework to explore the behavior of nonlinear oscillators \cite{genest2013dunkl, genest2014dunkl, quesne2023rationally,schulze2022bound}.

\subsection{Li\'{e}nard type - I: Isochronous oscillations}
We first discuss the quantum treatment for the various quadratic Li\'{e}nard type nonlinear oscillators possessing isochronous oscillations  (\ref{hamiltonian1-iso}), (\ref{hamiltonian2-iso}), (\ref{hamiltonian2-cubic}),  (\ref{hamiltonian2-negative}), and (\ref{hamiltonian2-power}) as studied in Ref. \cite{ruby2021quantum} and discussed in the earlier Section-II.

\subsubsection{Nonlinear oscillator- Exponential form}

The classical Hamiltonian which is associated with the nonlinear oscillator (\ref{hamiltonian1-iso}) having eight parameter Lie point symmetries in the exponential form is given by the expression,  
\begin{eqnarray}
\hspace{1cm}H_1 &=& \frac{1}{2\; \lambda^2}\;e^{-2\;\lambda\;x} p^2 + \frac{\omega^2_0}{2}\left(1 - e^{\lambda\;x}\right)^2,\nonumber \hspace{4.5cm}(\ref{hamiltonian1-iso})
\end{eqnarray}
where the mass term and the potential are expressed repectively as
\begin{equation}
\hspace{-0.5cm} \qquad \quad m(x) = \lambda^2 e^{2\;\lambda\;x}, \qquad V_1(x) = \frac{\omega^2_0}{2}\left(1 - e^{\lambda\;x}\right)^2. \label{mass1}
\end{equation}
The generalized Schr\"{o}dinger equation \eqref{seg_he} corresponding to the classical Hamiltonian (\ref{hamiltonian1-iso}) now takes the form
\begin{equation}
\psi'' - 2 \lambda \psi' +  \lambda^2 \left[A + \xi e^{2\;\lambda\; x} - \mu^2 (e^{2\;\lambda\;x} - 2 e^{3\lambda\;x} + e^{4\;\lambda\;x})\right] \psi = 0, \label{geq1}
\end{equation}
where the quantities $A,\; \xi, \mu$ are defined as 
\begin{eqnarray}
  A & = & - 4 {\overline{\alpha \gamma}} - (\bar{\gamma} - \bar{\alpha})^2 - 2 (\bar{\gamma} + \bar{\alpha}), \label{termA}\\
\xi & = & \frac{2\;E}{\hbar^2}, \\
\mu & = & \frac{\omega_0}{\hbar}. 
\end{eqnarray}
 
For positive values of $\lambda$ and for the following choice
\begin{eqnarray}
A &=& \frac{3}{4} \rightarrow 4 {\overline{\alpha\gamma}} + (\bar{\gamma} - \bar{\alpha})^2 + 2 (\bar{\gamma} + \bar{\alpha}) = -\frac{3}{4}, \label{avalue}
\end{eqnarray}
Eq. \eqref{geq1} yields the eigenfunctions 
\begin{equation}
\hspace{-0.5cm} \qquad \psi_n(x) = N_n\;\exp{\left(-\frac{\omega_0}{2\;\hbar}e^{2\;\lambda x} + \frac{\omega_0}{\hbar} e^{\lambda x}\right)}\; e^{\lambda x/2} \; 
H_n\left[\sqrt{\frac{\omega_0}{\hbar}}\;\left(e^{\lambda\;x}-1\right)\right], \quad -\infty < x < \infty, \label{eigen1}
\end{equation}
where $H_n(x)$ are Hermite polynomials, and $N_n, \; n = 0, 1, 2, 3, ...,$ represents the normalization constant. 
The energy eigenvalues $E_n$ and the normalization constant are given by
\begin{equation}
 E_n = \left(n + \frac{1}{2}\right)\;\hbar\;\omega_0,\qquad n = 0, 1, 2, 3, ...,  \label{en1}
\end{equation}
and 
\begin{eqnarray}
N_n = \left(e^{-\frac{\omega_0}{\hbar}}\;\frac{\sqrt{\frac{\omega_0}{\hbar}}\;\lambda}{\frac{\sqrt{ \pi}}{2}\;2^{n}\;n!\left(1+\text{erf}(a)\right)+a e^{-a^2}\mathcal{O}(a^2)}\right)^{1/2}, \quad a = \sqrt{\frac{\omega_0}{\hbar}},\label{nn1} 
\end{eqnarray}
where $\text{erf}(a)$ denotes the error function.

Similarly, we can derive the solution for negative values of $\lambda$ by substituting $\lambda = - |\lambda|$ in \eqref{eigen1}, yielding
\begin{equation}
\hspace{-0.5cm} \qquad \qquad \psi_n(x) = N_n\;\exp{\left(-\frac{\omega_0}{2\;\hbar}e^{-2\;|\lambda| x} + \frac{\omega_0}{\hbar} e^{-|\lambda| x}\right)}\; e^{-|\lambda| x/2} \; 
H_n\left[\sqrt{\frac{\omega_0}{\hbar}}\;\left(e^{-|\lambda|\;x}-1\right)\right],  -\infty < x < \infty,
\end{equation}
with energy eigenvalues, $E_n$, given as
\begin{equation}
 E_n = \left(n + \frac{1}{2}\right)\;\hbar\;\omega_0, \qquad n = 0, 1, 2, 3, ...,
 \label{en2}
\end{equation}
where $N_n, \; n = 0, 1, 2, 3, ...,$ represents the normalization constant, which can be derived as   
\begin{eqnarray}
N_n = \left(e^{-\frac{\omega_0}{\hbar}}\;\frac{\sqrt{\frac{\omega_0}{\hbar}}\;|\lambda|}{\frac{\sqrt{ \pi}}{2}\;2^{n}\;n!\left(1+\text{erf}(a)\right)+a\; e^{-a^2}\mathcal{O}(a^2)}\right)^{1/2}, \; a = \sqrt{\frac{\omega_0}{\hbar}}, \label{nn2} 
\end{eqnarray}
where $\text{erf}(a)$ denotes the error function.

The energy eigenvalues, which are presented in equations \eqref{en1} and \eqref{en2}, depend linearly on the quantum number $n$, and are not influenced by either the nonlinear parameter $\lambda$, or the ordering parameters.  Consequently, the isochronous nature of the classical system \eqref{ham1} is maintained in its corresponding quantum counterparts  \cite{ruby2021quantum}.

\subsubsection{Non-polynomial momentum dependent oscillator}

The non-polynomial nonlinear oscillator (\ref{ham2}) allows eight parameter Lie point symmetries, with
\begin{eqnarray}
\hspace{1cm}H_2 &=& \frac{1}{2}\;(1 + \lambda x)^4 p^2 + \frac{\omega^2_0}{2}\;\frac{x^2}{(1 + \lambda x)^2},\nonumber \hspace{5.3cm}(\ref{ham2})
\end{eqnarray}
where the mass term and the potential are identified as
\begin{equation}
m(x) = \frac{1}{(1 + \lambda x)^4}, \qquad V_2(x) = \frac{\omega^2_0}{2}\;\frac{x^2}{(1 + \lambda x)^2}. \label{mass2}
\end{equation}
The generalized Schr\"{o}dinger equation (\ref{seg_he}) now becomes 
\begin{equation}
\hspace{-0.5cm} \psi'' + \frac{4\;\lambda}{1 + \lambda x} \psi' +  \left[\frac{B\; \lambda^2}{(1 + \lambda x)^2} + \frac{2\;E}{\hbar^2\;(1 + \lambda x)^4} - \frac{\omega^2_0\;x^2}{\hbar^2\;(1 + \lambda\;x)^6}\right] \psi = 0, \label{geq2}
\end{equation}
where the term $B$ is defined as 
\begin{eqnarray}
  B & = & - 16 {\overline{\alpha \gamma}} - 4\;(\bar{\gamma} - \bar{\alpha})^2 - 6 (\bar{\gamma} + \bar{\alpha}). \label{termB}
\end{eqnarray}

For the case $\lambda > 0$, and by imposing the constraints on the ordering parameters, 
\begin{equation}
 \mbox{and}\qquad B = 2 \rightarrow 8 {\overline{\alpha \gamma}} + 2\;(\bar{\gamma} - \bar{\alpha})^2 + 3 (\bar{\gamma} + \bar{\alpha}) = -1,   \label{constraint2}
\end{equation} 
we can solve the corresponding Schr\"{o}dinger equation (\ref{geq2}) which leads to 
\begin{eqnarray}
\hspace{-2cm}\psi_n(x) = \left\{
\begin{array}{cc}
          N_n\;\frac{1}{(1 + \lambda x)}\;\exp{\left(-\frac{\omega_0\;x^2}{2\;\hbar\;(1 + \lambda x)^2}\right)}\; 
              H_n\left[\sqrt{\frac{\omega_0}{\hbar}}\;\left(\frac{x}{1 + \lambda x}\right)\right], 
							\quad -\frac{1}{\lambda} < x < \infty,\\
          \hspace{-0.5cm}0, \hspace{8.5cm}\quad x < -\frac{1}{\lambda}, 
\end{array}\right. 
\label{eigen3} 
\end{eqnarray}
with energy eigenvalues, $E_n$, as
\begin{equation}
E_n = \left(n + \frac{1}{2}\right)\;\hbar\;\omega_0,\qquad n = 0, 1, 2, 3, ..., \label{en3}
\end{equation}
where $N_n, \; n = 0, 1, 2, 3, ...,$ is the normalization constant obtained as 
\begin{eqnarray}
N_n = \left(\frac{\sqrt{\frac{\omega_0}{\hbar}}}{\frac{\sqrt{\pi}}{2}\;2^n\;n!\;\left(1+erf\left(c\right)\right)+c\;e^{-c^2}O(c^2)}\right)^{1/2}, \label{nn3} 
\end{eqnarray}
where $erf\left(\frac{1}{\lambda}\sqrt{\frac{\omega_0}{\hbar}}\right)$ is  the error function and ${\displaystyle c = \frac{1}{\lambda}\sqrt{\frac{\omega_0}{\hbar}}}$.  

For $\lambda < 0$, we can write down the eigenfunctions by substituting $\lambda = -|\lambda|$ in (\ref{eigen3}) as 
\begin{eqnarray}
\hspace{-2cm}\psi_n(x) = \left\{\begin{array}{cc}
\frac{N_n}{(1 -|\lambda| x)}\;\exp{\left(-\frac{\omega_0\;x^2}{2\;\hbar\;(1 - |\lambda| x)^2}\right)}\; 
H_n\left[\sqrt{\frac{\omega_0}{\hbar}}\;\left(\frac{x}{1 - |\lambda| x}\right)\right], \quad x \in \left(-\infty, 
\frac{1}{|\lambda|}\right),  \\
\hspace{-0.5cm}0,\hspace{8.5cm}  x > \frac{1}{|\lambda|}, \label{eigen3ne}
\end{array}\right. 
\end{eqnarray}
with energy eigenvalues, $E_n$, as
\begin{equation}
 E_n = \left(n + \frac{1}{2}\right)\;\hbar\;\omega_0, \qquad n = 0, 1, 2, 3, ...,
\end{equation}
where the normalization constant  $N_n, \; n = 0, 1, 2, 3, ...,$ is evaluated to be 
 
\begin{equation}
N_n = \left(\frac{\sqrt{\frac{\omega_0}{\hbar}}}{\frac{\sqrt{\pi}}{2}\;2^n\;n!\;\left(1+erf(c)\right)+c\;e^{-c^2}O(c^2)}\right)^{1/2}, \label{nn4} 
\end{equation} 
where ${\displaystyle c = \frac{1}{|\lambda|}\sqrt{\frac{\omega_0}{\hbar}}}$. 

The nonlinear oscillators (\ref{hamiltonian1-iso}) and (\ref{ham2}) governed by the generalized Schr\"{o}dinger equations (\ref{geq1}) and (\ref{geq2}) can be simplified to a bi-confluent equation  \cite{arscott1995heun}. We have demonstrated in this section that these equations admit a complete set of bounded solutions for specific parameter restrictions, namely (\ref{avalue}) and (\ref{termB}). It is possible to explore the quasi-exact solvable solutions for different parametric choices. The same approach can be applied to study all other isochronous systems, as explained in Ref. \cite{ruby2021quantum}.

\subsection{Li\'{e}nard type - I: Non-isochronous oscillations}
We consider the Hamiltonian for the nonlinear systems (\ref{ml}), (\ref{higg1d}), (\ref{nlo1d}) and (\ref{delta}) which show the non-isochronous oscillations at the classical level. 

\subsubsection{Mathews-Lakshmanan oscillator}

The generalized time-independent Schr\"{o}dinger equation for the Hermitian ordered quantum Hamiltonian (\ref{geo1d})  corresponding to the classical Mathews-Lakshmanan oscillator (\ref{ml}), 
\begin{equation}
H = \frac{(1 + \lambda x^2)}{2} p^2 +  \frac{\omega^2_0 x^2}{(1 + \lambda x^2)},  \nonumber 
\end{equation} 
takes the form, 
\begin{eqnarray}
\hspace{-0.5cm}\psi{''} + \frac{2\lambda x}{(1+ \lambda x^2)} \psi{'} + 
\left(\frac{A}{(1 + \lambda  x^2)}+ \frac{B}{(1 + \lambda x^2)^2} \right)\psi  = 0, \qquad  \left(' = \frac{d}{dx}\right)
\label{seg_mlo}
\end{eqnarray}
where 
\begin{eqnarray}
A &=& -4\lambda \bar{\alpha\gamma} -\lambda (\bar{\alpha}+ \bar{\gamma}) -\lambda (\bar{\gamma}- \bar{\alpha})^2 
+ \frac{2 E}{\hbar^2}-\frac{\omega^2_0}{\hbar^2 \lambda},
\label{A1-quantum}\\
B &=& 4\lambda \bar{\alpha\gamma} + \lambda (\bar{\gamma}- \bar{\alpha})^2 + \frac{\omega^2_0}{\hbar^2 \lambda}.  
 \label{B-quantum}
\end{eqnarray}
Eq. (\ref{seg_mlo}) has been solved exactly for both positive and negative values of $\lambda$, \cite{mathews1975quantum, karthiga2017quantum}.\\

{\bf (a) Negative values of $\lambda$:}

When $\lambda$ is negative, the dynamics of the system (\ref{ham-mlo}) is analyzed in the two regions, 
\begin{eqnarray}
\mbox{Region\;I}&:&\qquad -\frac{1}{\sqrt{|\lambda|}} \leq  x \geq \frac{1}{\sqrt{|\lambda|}}\\
\mbox{Region\;II}&:&\qquad  |x| > \frac{1}{\sqrt{|\lambda|}}, 
\end{eqnarray}
which are separated by the singular points. The solution to the Schr\"{o}dinger equation (\ref{seg_mlo}) in both regions is expressed as a combination of hypergeometric functions. The solutions obtained in both regions might be finite everywhere and continuous at their boundaries.

For $\lambda = -|\lambda|$, the  eigenfunctions are found out and expressed in terms of associated Legendre polynomials as   
\begin{equation}
\hspace{-2.5cm}\psi_n= \left\{\begin{array}{cc}
        N_n (1-|\lambda| x^2)^{\frac{\bar{\gamma}-\bar{\alpha}}{2}}\; P_{n+\mu}^{-\mu}(\sqrt{|\lambda|}x),\quad |x|<{|\lambda|}^{-\frac{1}{2}}\\
    \hspace{-1cm} 0, \quad |x|>|\lambda|^{-\frac{1}{2}},\,\, n =0, 1, 2, 3, ..., 
              \end{array}\right.\label{mo-sol2}
\end{equation}
where $\mu=\omega_0/\lambda \hbar$ and $N_n$ is the normalization constant, \cite{karthiga2017quantum}.  The corresponding energy spectrum of the later system is  
\begin{equation}
E_n = \left(n + \frac{1}{2}\right) \hbar \sqrt{\omega^2_0 + \hbar^2 \lambda^2 \left(4\bar{\alpha\gamma}+ (\bar{\gamma}-\bar{\alpha})^2\right)}+ \left(n^2 + n -\;\bar{\alpha} -\bar{\gamma}\right) \frac{\hbar^2 |\lambda|}{2}, n =0, 1, 2, 3, .... 
\label{mlo-energy2}
\end{equation}
Hence, we have an infinite set of bounded states restricted to the interior region.

The Schr\"{o}dinger equation (\ref{seg_mlo}) also admits the eigenfunctions with continuous energy eigen values, which vanish in the region-I and are non-zero in the region - II. Hence, for $|x| > \frac{1}{\sqrt{|\lambda|}}$, the generalized Schr\"{o}dinger equation admits the eigenfunctions, 
\begin{eqnarray}
\psi_{\rho}(x) = (1-|\lambda|x^2)^{-1/2}\;P^{-\mu}_{i\rho+\frac{1}{2}}(\sqrt{|\lambda|}x) \Theta(\pm\sqrt{|\lambda|}x - 1)
\end{eqnarray}
with continuous energy eigenvalues, 
\begin{eqnarray}
E_{\rho} = -\left(\rho^2 + \mu^2 + \frac{1}{4}+ \bar{\alpha}+\bar{\gamma}\right)\frac{\lambda\hbar^2}{2}, 
\end{eqnarray}
where $\Theta(\pm\sqrt{|\lambda|}x - 1)$ is known as Heaviside step function
$\Theta(y) = 1$ for $y > 0$ and $\Theta(y) = 0$ for $y < 0$.

{\bf (b) Positive values of $\lambda$:}

For positive values of $\lambda$, one can obtain a finite number of bound state eigenfunctions in the form
\begin{equation}
\hspace{-2.5cm}\psi_n= N_n (1+ \lambda x^2)^{\frac{\bar{\gamma}-\bar{\alpha}}{2}}\; Q_{\mu-n-1}^{\mu}(i\;\lambda^{\frac{1}{2}}x),\quad -\infty < x < \infty, 
\end{equation}
where $\mu=\omega_0/\lambda \hbar$ and $N_n$ is the normalization constant and $Q^{m}_{\nu}(z)$ are the associated Legendre functions of the second kind with $m$ and $\nu$ as parameters. And the energy eigenvalues are obtained as   
\begin{equation}
E_n = \left(n + \frac{1}{2}\right) \hbar \sqrt{\omega^2_0 + \hbar^2 \lambda^2 \left(4\bar{\alpha\gamma}+(\bar{\gamma}-\bar{\alpha})^2\right)} - \left(n^2 + n -\;\bar{\alpha} -\bar{\gamma}\right) \frac{\hbar^2 \lambda}{2}, \quad n = 0, 1, 2, ...N, 
\label{mlo-energy1}
\end{equation}
where $N$ is the largest allowed integer.  For the positive values of $\lambda$, one also has continuous energy states with $E_{\rho} = \left(\rho^2 + \mu^2 + \frac{1}{4}+ \bar{\alpha}+\bar{\gamma}\right)\frac{\hbar^2 |\lambda|}{2}$ which lie above the discrete energy states.

The Mathews-Lakshmanan oscillator has been studied in various generalized forms including exactly, quasi-exactly, and non-Hermitian versions \cite{midya2009generalized}. The rational extension of the quantum Mathews-Lakshmanan oscillator has also been constructed  to retain exact solvability, admitting a closed-form solutions that are expressed through Jacobi-type exceptional orthogonal polynomials \cite{schulze2013rational}.

%%%%%%%%%%%%%%%%%%%%%%%%%%%%%%%%%%%%%%%%%%%%%%%%
\subsubsection{Higgs oscillator}
For the Higgs oscillator, the classical Hamiltonian is of the form,
\begin{equation}
\hspace{1cm}H = (1 + k x^2)^2 \frac{p^2}{2} + \frac{\omega^2_0\; x^2}{2}.  \hspace{2cm}\label{non-ham-higgs}
\end{equation}
The generalized time-independent Schr\"{o}dinger equation (\ref{seg_he}) for the Hermitian ordered quantum Hamiltonian (\ref{non-ham-higgs}) becomes
\begin{eqnarray}
\hspace{-0.5cm} \;\hspace{-0.5cm}\psi{''} + \frac{4 k x}{(1+kx^2)} \psi{'} + 
\left(\frac{2 \eta_1 k -\frac{\omega^2_0 }{\hbar^2 k}}{(1 + k  x^2)}+ \frac{\frac{2 E}{\hbar^2}+ 4 \eta_2 k + \frac{\omega^2_0}{\hbar^2 k}}{(1 + k x^2)^2} \right)\psi  = 0, \qquad  \left(' = \frac{d}{dx}\right)
\label{seg_he1}
\end{eqnarray}
where 
\begin{eqnarray}
\eta_1 &=& 5(\bar{\alpha} + \bar{\gamma}) - 8 \left(\bar{\alpha\gamma} + \bar{\alpha} + \bar{\gamma}+\frac{1}{4}(\bar{\gamma}- \bar{\alpha})^2\right),
\label{eta1-quantum}\\
\eta_2 &=&-3(\bar{\alpha} + \bar{\gamma}) + 4 
 \left(\bar{\alpha\gamma} + \bar{\alpha} + \bar{\gamma}+\frac{1}{4}(\bar{\gamma}- \bar{\alpha})^2\right).  
 \label{eta2-quantum}
\end{eqnarray}

For positive values of $k$, the equation (\ref{seg_he1}) admits  the eigenfunctions,  
\begin{eqnarray}
\hspace{-0.5cm}\qquad \psi_n(x) = {\cal N}_n\;2^{\tilde{\mu}}\Gamma{(1+\tilde{\mu})} (1+ k\;x^2)^{-\frac{3}{4}}\;P^{-\tilde{\mu}}_{n+\tilde{\mu}}\left(\frac{\sqrt{k}\;x}{\sqrt{1+k\;x^2}}\right),\qquad k>0,  
\label{solnx}
\end{eqnarray}
where ${\displaystyle 0 < \frac{\sqrt{k}\;x}{\sqrt{1+k\;x^2}} < 1}$ provided that 
\begin{equation}
E_n = \left(n + \frac{1}{2}\right) \hbar \sqrt{\omega^2_0 + \hbar^2 k^2 \left(2\eta_1-\frac{9}{4}\right)}+ \left(n^2 + n + 2\;\bar{\alpha} + 2\bar{\gamma} + \frac{3}{2}\right) \frac{\hbar^2 k}{2}, 
\label{energy-higgs1}
\end{equation}
in which $\eta_1$ (vide Eq. (\ref{eta1-quantum})) is a function of the ordering parameters $\bar{\alpha}$ and $\bar{\gamma}$. 
Here, the normalization constant ${\cal N}_n$  is obtained as 
\begin{eqnarray}
 \;1 &=& \int^{\infty}_{-\infty} \psi^{*}_n(x) \psi_n(x) dx, \qquad k > 0 \\
 \mbox{so that}\nonumber\\ 
\; {\cal N}_n &=& \left(\frac{\sqrt{k}\;\Gamma(n+2\mu+1)\;(2n+2\mu+1)}{n!\;2}\right)^{1/2}.
\end{eqnarray}

Note that the presence of ordering parameters $\bar{\alpha}$ and $\bar{\gamma}$ in the energy eigenvalues (\ref{energy-higgs1}) indicates that different choices of ordering parameters results in different energy spectrum. 

Similarly, when $k < 0$ (say $k = -|k|$) the spatial region is divided into the following two regions: 
\begin{eqnarray}
\mbox{Region\;I}&:&\qquad -\frac{1}{\sqrt{|k|}} \leq  x \geq \frac{1}{\sqrt{|k|}}\\
\mbox{Region\;II}&:&\qquad  |x| > \frac{1}{\sqrt{|k|}}. 
\end{eqnarray}

We analyze the solvability of (\ref{seg_he1}) in accordance with the two regions. In Region I,  Eq. (\ref{seg_he1}) admits the admissible solutions as 
\begin{equation}
\psi_n(x) =  N_n \frac{\Gamma{\left(\frac{1}{2}-n+\tilde{\mu}\right)}}{\sqrt{(1-|k|\;x^2)}}\; P^{n-\tilde{\mu}+\frac{1}{2}}_{\tilde{\mu}-\frac{1}{2}}(\sqrt{|k|} x), 
\label{psikn}
\end{equation}
where $P^{m}_{\nu}(y)$ is known to be the associated Legendre polynomial with $m$ and $\nu$ as parameters. Here, $n = 0, 1, 2, ...N$ and the upper limit $N < 2 \tilde{\mu}-1$ and the normalization constant $N_n$ is evaluated to be 
\begin{eqnarray}
N_n = \left(\frac{\sqrt{|k|}\;\left(n-\tilde{\mu}+\frac{1}{2}\right)\;\Gamma{(2\tilde{\mu}-n)}}{\Gamma{(n+1)}\;
2^{\frac{n}{4}-\frac{\tilde{\mu}}{4}+\frac{1}{8}}}\right)^{1/2}. 
\end{eqnarray}
Energy eigenvalues can be evaluated for $k = -|k|$ as  
\begin{eqnarray}
E_n = \left(n + \frac{1}{2}\right)\hbar\omega_0\sqrt{1+ \frac{|k|^2 \hbar^2}{{\omega^2}_0}\;\left(\frac{9}{4}-2 \eta_1\right)} - \left(n^2 + n +2 \bar{\alpha} +  2 \bar{\gamma}+\frac{3}{2}\right)\frac{\hbar^2 |k|}{2}, \nonumber \\ 
\hspace{6cm}\qquad n = 0, 1, 2, 3, ...N. 
\end{eqnarray}
The energy eigenvalues $E_n$ are discrete and finite in number. And above the upper limit of $n$, that is $N$, the energy eigenvalue is continuous. 

For the region-II, that is $|x| > \frac{1}{\sqrt{|k|}}$, the eigenfunctions are obtained as 
\begin{eqnarray}
\psi_{\rho}(x) = (1-|k|x^2)^{-1/2}\;P^{-i\rho}_{\tilde{\mu}-\frac{1}{2}}(\sqrt{|k|}x) \Theta(\pm\sqrt{|k|}x - 1), \label{psik-2}
\end{eqnarray}
with energy eigenvalues, 
\begin{eqnarray}
E_{\rho} = \left(\frac{\rho^2 + \mu^2 + 1}{2}+2\;\eta_2\right)\hbar^2 |k|, 
\end{eqnarray}
where $\Theta(\pm\sqrt{|k|}x - 1)$ is the Heaviside step function. 

\subsubsection{$k$-dependent non-polynomial oscillator}
The classical Hamiltonian for the system (\ref{nlo1d}) is 
\begin{equation}
H_2 = (1 + k x^2)^2 \frac{p^2}{2}+ \frac{\omega^2_0 x^2}{2 (1 + k x^2)^2}.  \label{non-ham-nlo}
\end{equation}

It has been observed that the non-Hermitian ordered form of the Hamiltonian (\ref{non-ham-nlo}) admits well defined eigenfunctions. The associated time-independent one dimensional generalized Schr\"{o}dinger equation is
\begin{eqnarray}
\hspace{-0.5cm}\Phi{''} + \frac{4 k x}{(1+kx^2)}(1+\bar{\alpha}-\bar{\gamma}) \Phi{'} + 
\left(\frac{4 \sigma_1 k}{(1 + k  x^2)}+ \frac{\frac{2 E}{\hbar^2}+ 4 \sigma_2 k}{(1 + k x^2)^2} - \frac{\omega_0^2 x^2}{\hbar^2\;(1 + k x^2)^4} \right)\Phi  = 0, 
\label{seg_hen}
\end{eqnarray}
where 
\begin{eqnarray}
\sigma_1 &=& -4 \bar{\alpha\gamma} - 3 \bar{\gamma}, 
\label{sigma1}\\
\sigma_2 &=& 4 \bar{\alpha\gamma} + 2 \bar{\gamma}. 
\label{sigma2}
\end{eqnarray}
Eq. (\ref{seg_hen})  is of the form of confluent Heun equation which can be solved using Bethe ansatz method \cite{zhang2012exact,quesne2017families}. The eigenfunctions, $\Phi_n(x)$, are obtained as  
\begin{eqnarray}
\hspace{-0.5cm} \Phi^{(l)}_n(x) =N^{(l)}_n \exp{\left(-\frac{\omega_0\;x^2}{2\hbar(1+k x^2)}\right)} (1 + k x^2)^{n+\bar{\gamma}-\bar{\alpha}} (k x^2)^{l}\;\; \Pi^{n}_{i = 0} \left(\frac{k x^2}{1 + k x^2} - z_i\right)
\label{non-psinx}
\end{eqnarray}
and the energy eigenvalues are 
\begin{eqnarray}
\hspace{-0.5cm} E^{(l)}_n =\left(2n +2l+ \frac{1}{2}\right)\hbar \omega_0 +\left(\sigma_1+ \bar{\gamma}-(\bar{\gamma}-\bar{\alpha})(\bar{\gamma}-\bar{\alpha}-1)-\mu\sum^{n}_{i = 1} z_i  \right)\;2\hbar^2 k \label{non-enf}
\end{eqnarray}
with 
\begin{equation}
\sigma_1 = \mu \sum^{n}_{i} z^2_i + (2- \mu)\sum^{n}_{i = 1} z_i-n^2 -2l+(\bar{\gamma}-\bar{\alpha})\left(\bar{\gamma}-\bar{\alpha}-\frac{3}{2}\right),  
\label{sigman1-non}
\end{equation} 
and the roots $z_i$'s, $i = 1, 2, 3, ... N, $ satisfy the equation, 
\begin{equation}
\sum^n_{j\neq i} \frac{2}{z_i - z_j}=- \frac{\mu z_i^3 + \left(-\mu + 2 n-2l-\frac{1}{2}\right) z_i + \left(2\mu - 2n\right) z^2_i +2 l + \frac{1}{2}}{z_i(1-z_i)^2}.\label{bethe-ansatz1}
\end{equation}
Here it is also noted that the term $\sigma_1$ containing the ordering parameters  is related with the quantum number $n$ given in (\ref{sigman1-non}).

\subsection{Li\'{e}nard type - I: Localized Solutions}
As the mass $m(x)$ of the system (\ref{delta}) is singular at $x=0$, we use the single term of the general ordered form of the Hamiltonian as
\begin{eqnarray}
\hat{H} = \frac{1}{2} m^{\alpha_1} \hat{p} m^{\beta_1} \hat{p} m^{\gamma_1} + V(x), \qquad \alpha_1+\beta_1+\gamma_1 = -1.  \label{single}
\end{eqnarray}

The time-independent Schr\"{o}dinger equation for the non-Hermitian ordered Hamiltonian (\ref{single}), $\hat{H}\psi = E \psi,$ can be written as   
\begin{eqnarray}
\hspace{-0.5cm}\quad \psi{''} +\left({\gamma}_1-{\alpha}_1-1\right) \frac{m{'}}{m} \psi{'} + 
\left({\gamma}_1\;\frac{m{''}}{m}-\left({\alpha_1\gamma_1}+2{\gamma_1}\right)  \frac{m{'^2}}{m^2}\right) \psi + \frac{2 m}{\hbar^2}\left(E -V(x)\right)\psi  = 0,  \nonumber\\
\label{SE1}
\end{eqnarray}
where ${\displaystyle ' = \frac{d}{dx}}$. 

In general, Eq. (\ref{SE1}) has been solved for two regions (i) $x \in (-\infty, 0)$ and (ii) $x \in (0, \infty)$, 
\cite{ruby2022quantum}. By using the parity nature of the eigenfunctions obtained in both the regions, one can obtain bound state solutions for (\ref{SE1}) in the region-I, $x \in (0, \infty)$, and for the region-II, $x \in (-\infty, 0)$,    
\begin{eqnarray}
\hspace{0.1cm} \psi^{(+)}_{n}(x) &=&  C J_{n}\left(\frac{2\sqrt{E}}{\hbar\; x}\right),  \quad x \in (0, \infty) \quad n=1,2, 3, ....\label{sol_p}\\
\hspace{0.1cm} \psi^{(-)}_{n}(x) &=&  C (-1)^n J_{n}\left(\frac{2\sqrt{E}}{\hbar\; |x|}\right),  \quad x \in (-\infty, 0) \quad n=1,2, 3, ....\label{sol_m}
\end{eqnarray}
In the above, $J_n(a x)$ is the Bessel function of the first kind. 
Here, the energy eigenvalues are continuous, while the coupling parameter $\lambda$ is quantized as 
\begin{eqnarray}
\lambda = \left(n^2 - \left(2{\alpha}_1+2{\gamma}_1+\frac{3}{2}\right)^2\right)\frac{\hbar^2}{4}, \quad n = 1, 2, 3, ... .  \label{lambda-n}
\end{eqnarray}

It is also observed that one could possibly obtain the normalized eigenfunctions with the corresponding eigenvalues by restricting the motion of the particle around a point near to the origin \cite{ruby2021classical}. 

\subsection{Isotonic generalization of non-isochronous nonlinear oscillators:}
The classical studies carried out in the earlier sections reveal that the Mathews-Lakshmanan oscillator and the Higgs oscillator continue to exhibit periodic oscillations with the inclusion of an isotonic term.  The quantum solutions for these oscillators are of particular interest. Moreover, studying the effects of isotonic terms on different types of potentials can provide valuable insights into the behavior of quantum systems with non-polynomial oscillators. For example, while solving the  
$k$-dependent non-polynomial oscillator (\ref{nlo1d-iso}) quantum mechanically, the inclusion of the isotonic term adds complexity to its classical solvability. Therefore, investigating the isotonic generalization of  the quantum version of the nonlinear system may lead to quasi-exactly solvable solutions. But in the case of $\delta$-type potential (\ref{delta-iso}), the addition of isotonic term removes the complexity in its quantum counterpart.  This indicates that the isotonic term plays a crucial role in determining the solvability of the system.

\subsubsection{Isotonic generalization of Mathews-Lakshmanan oscillator}

The classical Hamiltonian for the isotonic Mathews-Lakshmanan oscillator is taken to be of the form, 
\begin{equation}
H = \frac{(1- \lambda x^2)\;p^2}{2} + \frac{\omega^2_0\;x^2}{2\;(1 - \lambda x^2)} + \frac{g}{2\;x^2}, \quad \lambda > 0, \label{ham-ml-iso}
\end{equation}
where $\omega_0, \lambda$ and $g$ are system parameters. The generalized Schr\"{o}dinger equation (\ref{seg_he}) for the hermitian ordered form of the Hamiltonian (\ref{ham-ml-iso}) has been solved and it leads to the eigenfunctions,        
\begin{equation}
\hspace{-0.5cm}\psi_n= \left\{\begin{array}{cc}
        N_n\;(\lambda x^2)^{\frac{1}{4}+\frac{\tilde{g}}{2}} (1 - \lambda x^2)^{\frac{\tilde{\mu}}{2}}\; P^{(-\tilde{\mu}, \tilde{g})}_{n}(1-2 \lambda x^2),\quad |x|<\lambda^{-\frac{1}{2}}\\
    \hspace{-1cm} 0, \quad |x|> \lambda^{-\frac{1}{2}},\,\, n =0, 1, 2, 3, ..., 
              \end{array}\right.
\end{equation}
and  the energy spectrum
\begin{eqnarray}
E_n &=& \left(2 n + \tilde{g} + 1\right)\hbar \sqrt{\omega^2_0 + g \lambda^2+\hbar^2 \lambda^2 \left(4\bar{\alpha\gamma}+(\bar{\gamma}-\bar{\alpha})^2\right)}\nonumber \\
&+& \left(\left(2 n + \tilde{g} + 1\right)^2 -\frac{1}{4}-(\bar{\alpha} +\bar{\gamma})\right)2\hbar^2 \lambda. 
\end{eqnarray}
Here  $ P^{(-\tilde{\mu}, \tilde{g})}_{n}(z)$ is known as the Jacobi polynomial and the other parameters are defined as  
\begin{eqnarray}
\tilde{\mu}&=&\sqrt{\frac{\omega^2_0}{\lambda^2 \hbar^2} + \frac{g}{\hbar^2}+4 \bar{\alpha \gamma} + (\bar{\gamma}-\bar{\alpha})^2},
\quad 
\tilde{g} = \sqrt{\frac{1}{4}+\frac{g}{\hbar^2}}. 
\end{eqnarray}

Recently, the Dunkl version of the quantum Hamiltonian has been constructed for the isotonic Mathews-Lakshmanan oscillator, \cite{schulze2022bound}. It is observed that the corresponding generalized Schr\"{o}dinger equation admits bound states for a very restrictive set of parameters, only in the absence of isotonic term. It is an interesting variant of the conventional isotonic Mathews-Lakshmanan oscillator (\ref{ham-ml-iso}).

\subsubsection{Isotonic generalization of Higgs oscillator}
The generalized Schr\"{o}dinger equation (\ref{seg_he}) for the Higgs oscillator with isotonic term, 
\begin{equation}
H = (1 + k x^2)\frac{p^2}{2}+\frac{\omega^2_0}{2} x^2+\frac{g}{2\;x^2}, \label{iso-higgs}
\end{equation}
where $k, \omega_0$ and $g$ are system parameters, can be written as 
\begin{eqnarray}
\hspace{-0.5cm}\psi{''} + \frac{4 k x}{(1+kx^2)} \psi{'} + 
\left(\frac{2 \eta_1 k -\frac{\omega^2_0 }{\hbar^2 k} + \frac{g k}{\hbar^2}}{(1 + k  x^2)}+ \frac{\frac{2 E}{\hbar^2}+ 4 \eta_2 k + \frac{\omega^2_0}{\hbar^2 k}+\frac{g k}{\hbar^2}}{(1 + k x^2)^2} -\frac{g}{\hbar^2\;x^2}\right)\psi  = 0, \;  \left(' = \frac{d}{dx}\right) \nonumber \\
\label{seg-he1-iso}
\end{eqnarray}
where 
\begin{eqnarray}
\eta_1 &=& 5(\bar{\alpha} + \bar{\gamma}) - 8 \left(\bar{\alpha\gamma} + \bar{\alpha} + \bar{\gamma}+\frac{1}{4}(\bar{\gamma}- \bar{\alpha})^2\right),
\label{eta1}\\
\eta_2 &=&-3(\bar{\alpha} + \bar{\gamma}) + 4 
 \left(\bar{\alpha\gamma} + \bar{\alpha} + \bar{\gamma}+\frac{1}{4}(\bar{\gamma}- \bar{\alpha})^2\right).  
 \label{eta2}
\end{eqnarray}

On using simple transformations, Eq. (\ref{seg-he1-iso}) can be solved for positive values of $k$ and it admits eigenfunctions, 
\begin{eqnarray}
\psi_n(x) = N_n \frac{(kx^2)^{\frac{1}{4}+\frac{\tilde{g}}{4}}}{(1+ kx^2)^{1+\frac{\tilde{\mu}}{4}+\frac{\tilde{g}}{4}}}\;P_n^{\left(\frac{\tilde{\mu}}{2}, \frac{\tilde{g}}{2}\right)}\left(\frac{1-kx^2}{1+kx^2}\right), \label{seg-sol-higgs}
\end{eqnarray}
where
\begin{eqnarray}
\tilde{\mu} = \sqrt{4 \mu^2 - 9 - 8 \eta_1}, \qquad \qquad \tilde{g} = \sqrt{1 + \frac{4 g}{\hbar^2}}.
\end{eqnarray}
And the energy eigenvalues are 
\begin{eqnarray}
E_n = \left(n+\frac{1}{2}\right) (\tilde{\mu} + \tilde{g}) + \left(n^2 + n + \eta_2-\frac{\eta_1-1}{2}\right)2\hbar^2 k.
\end{eqnarray}

\subsubsection{$k$-dependent non-polynomial oscillator with isotonic term}
Consider the classical Hamiltonian for the system (\ref{nlo1d}) with an addition of inverse square term as  
\begin{equation}
H_2 = (1 + k x^2)^2 \frac{p^2}{2}+ \frac{\omega^2_0 x^2}{2 (1 + k x^2)^2} + \frac{g}{2\;x^2}.  \label{ham-nlo-iso}
\end{equation}

The time-independent one dimensional generalized Schr\"{o}dinger equation associated with the Hermitian ordered form of the Hamiltonian (\ref{seg_he}) is
\begin{eqnarray}
\hspace{-0.5cm}\;\psi{''} + \frac{4 k x}{(1+kx^2)}\psi{'} + 
\left(\frac{4 \eta_1 k}{(1 + k  x^2)}+ \frac{\frac{2 E}{\hbar^2}-\frac{2 g k}{\hbar^2}+ 4 \eta_2 k}{(1 + k x^2)^2} - \frac{\omega_0^2 x^2}{\hbar^2\;(1 + k x^2)^4} -\frac{g}{\hbar^2\;x^2} \right)\psi  = 0, \nonumber \\
\label{her-iso}
\end{eqnarray}
where $\eta_1$ and $\eta_2$ are defined in the equations (\ref{eta1}) and (\ref{eta2}). 

On using the transformation, $\psi(x) = x^{2 l}(1+k x^2)^d \exp{\left(-\frac{\mu\; x^2}{2\;(1 + k x^2)}\right)}S(x)$, we can transform the equation (\ref{her-iso}) to be, 
\begin{eqnarray}
\hspace{-0.5cm} S{''} + \left[\frac{4 l}{x} + \frac{4 k (d+1)\;x}{(1+kx^2)-\frac{2\mu k x}{(1+k x^2)}}\right]S{'} + 
\left[\frac{2l(2l-1)-g/{\hbar^2}}{x^2}+\frac{4 \bar{\eta} k}{(1 + k  x^2)} + \frac{4\bar{\xi} k}{(1 + k x^2)^2}\right. \nonumber \\
\left.+\frac{4 \mu d k}{(1+kx^2)^3} \right]S= 0, 
\label{her-non-se}
\end{eqnarray}
with 
\begin{equation}
\mu = \frac{\omega_0}{\hbar\;k}, 
\end{equation}
and $\bar{\eta}$ and $\bar{\xi}$ take the forms, 
\begin{eqnarray}
\bar{\eta} &= &\frac{\eta_1}{2} + d\left(d + \frac{3}{2}\right) + 2l (d + 1), \label{eta-iso}\\
\bar{\xi} &=& \frac{E}{2\hbar^2 k} +\eta_2-d(d+1)-\frac{g}{2\hbar^2}-\mu(l+d)-\frac{\mu}{4}. \label{xi-iso}
\end{eqnarray}

On further using the transformation, $z = \frac{k x^2}{1 + k x^2}$, we can reduce the equation (\ref{her-non-se}) to the form of confluent Heun equation, 
\begin{eqnarray}
\hspace{-0.5cm} z (1 - z)^2\frac{d^2 S}{dy^2} +\left[2 l +\frac{1}{2} + \left(2d-\mu-2l-\frac{1}{2}\right) z +2(\mu-d)z^2-\mu z^3\right]\frac{dS}{dy} \nonumber \\
+\left[\bar{\xi} + \bar{\eta}+\mu d-(\bar{\xi}+2\mu d)z +\mu d z^2\right]S(y) = 0. \label{bethe-k-se}
\end{eqnarray}
Eq. (\ref{seg_hen}) can also be solved using Bethe ansatz method and the  eigenfunctions, $\psi_n(x)$, are obtained as  
\begin{eqnarray}
\hspace{-0.5cm} \qquad \psi_n(x) =N_n \exp{\left(-\frac{\omega_0\;x^2}{2\hbar (1+k x^2)}\right)}\;x^{2 l} (1 + k x^2)^{d} \; \Pi^{n}_{i = 0} \left(\frac{k x^2}{1 + k x^2} - z_i\right)
\label{psinx-non}
\end{eqnarray}
and the energy eigenvalues are 
\begin{eqnarray}
E^{(l)}_n =\left(2n +2l+ \frac{1}{2}\right)\hbar \omega_0 +\left(-\mu\sum^{n}_{i = 1} z_i - 2 \eta_2 +g k/2\right)\;2\hbar^2 k \label{enf}
\end{eqnarray}
with 
\begin{equation}
d = n, \; 2l(2l-1) = g/\hbar^2 > 0, \; \mu \sum^{n}_{i} z^2_i + (2- \mu)\sum^{n}_{i = 1} z_i-n(n-2l+3/2)+\bar{\eta} = 0.
\label{sigman1}
\end{equation} 
Here the isotonic term $g$ is related with $l$ and  the term $\bar{\eta}$ containing ordering parameters, $\alpha_i +\beta_i +\gamma_i = -1,\;i =1, 2, 3, ... N$, which are considered to be arbitrary. 

The eigenfunctions (\ref{psinx-non}) are not bounded when  $x \rightarrow \pm \infty$ for $k>0$, whereas for $k < 0$, that is $k = - |k|$, they are bounded within  $\frac{-1}{\sqrt{|k|}}< x < \frac{1}{\sqrt{|k|}}$. As the results obtained in  the case of $k$-dependent nonpolynomial oscillator show that the corresponding non-hermitian ordered form of the Hamiltonian produces bounded eigenfunctions, we evaluate the eigenfunctions corresponding to the non-hermitian ordered form of the Hamiltonian (\ref{ham-nlo-iso}) via the general relation, $\phi = m^{\frac{\bar{\alpha}-\bar{\gamma}}{2}} \psi$. Hence the eigenfunctions for the non-hermitian ordered Hamiltonian can be written as 
\begin{eqnarray}
\hspace{-0.5cm}\phi_n(x) =N_n \exp{\left(-\frac{\omega_0\;x^2}{2\hbar (1+k x^2)}\right)}\;x^{2 l} (1 + k x^2)^{n+\bar{\gamma}-\bar{\alpha}} \; \Pi^{n}_{i = 0} \left(\frac{k x^2}{1 + k x^2} - z_i\right)^n, 
\label{psinx}
\end{eqnarray}
which are bounded by fixing $N+\bar{\gamma}-\bar{\alpha} < 0$ and $N$ is the upper limit of the allowed energy levels. 

\subsection{$\delta$-type potential with isotonic term}
The isotonic generalized form of the classical Hamiltonian (\ref{hamiltonian}) reads as
\begin{equation}
H = \frac{x^4\;p^2}{4} + \lambda x^2 +\frac{g}{2\, x^2}, \qquad g > 0,  
\label{hamiltonian-iso}
\end{equation}
where $g$ is arbitrary and the mass $m(x) = \frac{1}{x^4}$ is singular, we consider the single ordered Hamiltonian operator (\ref{single}). 

The associated time-independent Schr\"{o}dinger equation for the non-Hermitian ordered Hamiltonian (\ref{single}), $\hat{H}\psi = E \psi,$ can be written as   
\begin{eqnarray}
\hspace{-0.5cm}\quad \psi{''} +\left({1-\gamma}_1+{\alpha}_1\right) \frac{2}{x} \psi{'} + 
\left[-\frac{4\lambda}{\hbar^2}+\frac{\frac{4 E}{\hbar^2}-2\gamma_1-4\alpha_1\gamma_1}{x^2}-\frac{2 g}{\hbar^2 x^4}\right]\psi  = 0, \nonumber \\
\label{se1-iso}
\end{eqnarray}
where ${\displaystyle ' = \frac{d}{dx}}$. 

By using the transformations, $\psi(x) = x^{(-l)} e^{-\mu/4 x^2} S(x)$ and $y = \frac{1}{\sqrt{2}\;x}$, we can reduce the equation (\ref{se1-iso}) to be the form 
\begin{eqnarray}
\hspace{-0.5cm} y S''(y) + \left[1+2l - (\alpha_1-\gamma_1+1) y +\mu y^2\right] S'(y) + \left[-2l(\alpha_1-\gamma_1+1)+(2\mu l + \xi + 2\mu) y \right. \nonumber \\
\left.- 2\mu(\alpha_1-\gamma_1 +1)y^2\right]S = 0 \qquad \label{delta-che}
\end{eqnarray}
with 
\begin{eqnarray}
\mu &=& \frac{2}{\hbar}\;\sqrt{g}, \\
l\hbar &=& \sqrt{\frac{\lambda}{2}}, \label{lvalue}\\
\xi &=& \frac{4E}{\hbar^2} -  2\gamma_1 - 4\alpha_1\gamma_1. 
\end{eqnarray}

Then, we use the Bethe anstaz \cite{zhang2012exact} method to solve equation (\ref{delta-che}) which yields the energy eigenvalues as 
\begin{equation}
E_n = (n + 2)\frac{\hbar\sqrt{g}}{2} + \sqrt{\frac{g \lambda}{2}} + \left(\alpha_1\gamma_1 + \frac{\gamma_1}{2}\right)\hbar^2, 
\end{equation}
and the eigenfunctions are obtained as 
\begin{equation}
\psi_n(x) = e^{-\frac{\mu}{4x^2}}\left(\frac{1}{\sqrt{2}x}\right)^l\;\Pi^n_{i=1}\left(\frac{1}{\sqrt{2\;x}} - y_i\right)^n, n =0, 1, 2, 3, ... ,\label{delta-iso-eig}
\end{equation}
where the roots, $y_i, \; i = 1, 2, 3, ...n$, satisfy the relation 
\begin{equation}
\mu \sum^n_{i=1} y_i + (n+2l)(\gamma_1-\alpha_1+1) = 0. \label{root-eqn}
\end{equation}
The eigenfunctions (\ref{delta-iso-eig}) are bounded at the region $x = 0$ as the exponential term becomes zero at the origin. Similarly when $x - \pm \infty$, $\psi_n(x)$ becomes zero provided $l > 0$. 

\subsection{{\label{qhigg3d} Quantum solvability of the three dimensional Mathews-Lakshmanan oscillator}}
We use polar coordinates to define the Hamiltonian for the three-dimensional Mathews-Lakshmanan oscillator in order to comprehend the corresponding quantum dynamics.  The form of the Hamiltonian that corresponds to the Lagrangian (\ref{gasiroq}) is as follows, 
\begin{eqnarray}
H = \frac{1}{2}\left[\frac{\dot{r}^2}{(1 + \lambda r^2)^2} + r^2\dot{\theta}^2+ r^2\sin^2\theta \dot{\phi}^2 + \frac{\omega^2_0 r^2}{(1+\lambda r^2)} \right]. \label{hamD3-mlo}
\end{eqnarray}
where the radial coordinate is represented by $'r'$, while the polar and azimuthal angles are denoted by $\theta$ and $\phi$, respectively. The associated conjugate momenta are also calculated as 
\begin{equation}
p_r = \frac{\dot{r}}{(1 + \lambda r^2)}, \quad p_{\theta} = r^2 \dot{\theta}, \quad {\text and}\quad  p_{\phi} = r^2\;\sin^2{\theta}\dot{\phi}.   
\label{momD3-mlo}
\end{equation}
The Hamiltonian (\ref{hamD3-mlo}) can be re-expressed in terms of momentum coordinates as
\begin{eqnarray}
H = \frac{1}{2}\left[(1 + \lambda r^2) p_r^2 +\frac{1}{r^2}\left(p^2_{\theta} + \frac{p^2_{\phi}}{\sin^2{\theta}}\right) + 
\frac{\omega^2_0 r^2}{(1 + \lambda r^2)} \right]. \label{hamD3-mlo-momentum}
\end{eqnarray}

The quantum version of the three-dimensional Mathews-Lakshmanan oscillator (\ref{hamD3-mlo-momentum}) has been solved using a symmetric ordering form for the Hamiltonian operator \cite{lakshmanan1975quantum}. We are now interested in exploring how the solvability of the system changes with different ordering forms, as this will provide insight into the significance of orderings in quantization. To achieve this, we consider the three-dimensional version of the general ordered form (\ref{geo1d}), 

\begin{eqnarray}
\hat{H} = \frac{1}{2}\left[\sum^N_{i = 1} w_i m^{\alpha_i} \hat{p}_r m^{\beta_i} \hat{p}_r m^{\gamma_i}+\frac{1}{r^2}\left(p^2_{\theta} + \frac{p^2_{\phi}}{\sin^2{\theta}}\right) + \omega^2_0 r^2 \right],
\label{geo}
\end{eqnarray}
where  $N$ is an arbitrary positive integer, and the ordering parameters should satisfy the constraint $\alpha_i +\beta_i +\gamma_i = -1,\;i =1, 2, 3, ... N$ and $w_i$'s  are real weights which are summed to be $1$. Here $m$ is the position dependent mass associated with the Hamiltonian (\ref{hamD3-mlo-momentum}). 

The  Hermitian Hamiltonian  (\ref{hermitian})  for a potential and a mass function corresponding to polar coordinates can be expressed as 
\begin{eqnarray}
\hspace{-0.5cm} \hat{H}_{her} &= &\frac{1}{2}\left(\frac{1}{m}\hat{p}^2_r - i \hbar \frac{d}{d\;r}\left(\frac{1}{m}\right)\hat{p}_r\right) + \frac{\hbar^2}{2}\left[\left(\frac{\bar{\alpha}+\bar{\gamma}}{2}\right) \frac{d^2}{d r^2}\left(\frac{1}{m}\right) + \left(\overline{\alpha\gamma}+ \frac{1}{4} (\bar{\gamma} - \bar{\alpha})^2\right)  \left(\frac{d}{dr}\left(\frac{1}{m}\right)\right)^2 m\right.\nonumber\\
\hspace{-0.5cm} & &\left. + \frac{1}{\;r^2}\left(\hat{p}^2_{\theta} + \frac{\hat{p}^2_{\phi}}{\sin^2{\theta}}\right)\right] + V(r),  
\label{geham}
\end{eqnarray}
where ${\displaystyle \hat{p}_r = - i \hbar \left(\frac{\partial}{\partial r} + \frac{1}{r}\right)},\quad {\displaystyle \hat{p}_{\theta} = - i \hbar \left(\frac{1}{\sin{\theta}} \frac{\partial}{\partial \theta}\right)}$ and ${\displaystyle \hat{p}_{\phi} = -i\hbar \frac{\partial}{\partial \phi}}$. 

The generalized Schr\"{o}dinger equation corresponding to the Hermitian ordered form (\ref{geham}) of the Hamiltonian (\ref{geo})  can be written as
\begin{eqnarray} 
\hspace{-0.5cm} \Psi''+ \left(\frac{2}{r}-\frac{m'}{m}\right)\Psi'+ \left[\left(\frac{\bar{\alpha}+\bar{\gamma}}{2}\right) \left(\frac{m''}{m}\right) - \left(\overline{\alpha\gamma}+\bar{\alpha}+\bar{\gamma}+ \frac{1}{4} (\bar{\gamma} - \bar{\alpha})^2\right)  \left(\frac{m'^2}{m^2}\right) \right.\nonumber \\
\hspace{-2cm}\left.-\frac{\hat{L}^2}{r^2\;}+ \frac{2\;m}{\hbar^2}\left(E - V(r)\right)\right]\Psi = 0,  
\label{geham_heo3d}
\end{eqnarray}
where the square of the angular momentum operator  $\hat{L}^2$ is
\begin{equation}
\hat{L}^2 =-\hbar^2\left(\frac{1}{\sin{\theta}}\frac{\partial}{\partial \theta}\left(\frac{1}{\sin{\theta}}\frac{\partial}{\partial \theta}\right)+\frac{1}{\sin^2{\theta}}\frac{\partial^2}{\partial \phi^2}\right),
\end{equation}
and $' = \frac{\partial}{\partial r}$. 
Considering the wave function to be  of the form,
${\displaystyle \Psi(r, \theta, \phi) = \frac{\chi(r)}{r} Y(\theta, \phi)}$, with a spherically symmetric factor 
${\displaystyle \frac{\chi(r)}{r}}$ with $m(r) = \frac{1}{(1 + \lambda r^2)}$ and potential $V(r) = \frac{\omega^2_0 r^2}{(1 + \lambda r^2)}$ and a generalized angular function $Y(\theta, \phi)$, Eq. (\ref{geham_heo3d}) can be separated as follows: 
\begin{eqnarray}
\hspace{-0.5cm}\;\;\mbox{Angular part:}&\;\;\;\;\;\;\;L^2 Y(\theta, \phi) = \hbar^2\;l(l+1)\; Y(\theta, \phi),& \label{a3dim-mlo}\\
\hspace{-0.5cm}\;\;\mbox{Radial part:}&\;\;{\displaystyle \chi''+\frac{2\;\lambda r}{(1+\lambda\;r^2)}\chi'+ \left[\frac{A}{(1+\lambda\;r^2)} + \frac{B}{(1+\lambda\;r^2)^2} - \frac{l(l+1)}{r^2}\right]\chi = 0,} &
\label{r3dim-mlo}
\end{eqnarray}
where $Y_{l, m}(\theta, \phi) = \sqrt{\frac{(l - m)!}{(l + m)!}}\; e^{i m \phi}\; P^m_l(\cos{\theta})$ are spherical harmonics. Here $P^m_l$ is the associated Legendre polynomial, $l$ is the angular momentum quantum 
number and $m$ is the magnetic quantum number. For every value of $l$, $m = -l, -(l-1), ... 0, ... l-1, l$, that is $m$ takes $2 l + 1$ values. Here 
\begin{eqnarray}
A &=& -2\lambda-4\bar{\alpha\gamma}-\lambda(\bar{\gamma}+\bar{\alpha})-\lambda (\bar{\gamma}-\bar{\alpha})^2 -\lambda\;\mu^2 + \lambda l(l+1)+\frac{2 E}{\hbar^2},\label{energy-mlo-3d1}\\
B &=& 4\bar{\alpha\gamma}+\lambda (\bar{\gamma}-\bar{\alpha})^2 + \lambda\;\mu^2,\\
\mu &=& \frac{\omega_0}{\hbar\;k}.\label{muh}
\end{eqnarray}

\subsubsection*{\bf (a) Positive values of $\lambda$}
When $\lambda > 0$, $r$ spreads over from $0$ to $\infty$. On using a set of transformations on (\ref{a3dim-mlo}) and (\ref{r3dim-mlo}), we can solve the Schr\"{o}dinger equation (\ref{geham_heo3d}). Let us consider the transformation, 
\begin{eqnarray}
\sqrt{\lambda}\;r =\frac{\sqrt{z}}{\sqrt{1-z}}. 
\label{rtrans}
\end{eqnarray}
Then Eq. (\ref{r3dim-mlo}) gets reduced to  
\begin{equation}
z(1 -z)\chi''(z) + \left(\frac{1}{2}-z\right) \chi'(z) + \left[\frac{B}{4\;\lambda} + \frac{\frac{A}{4\lambda}-\frac{l(l+1)}{4}}{\;(1 - z)}- \frac{l(l+1)}{4\;z}\right] \chi(z) = 0.   
\label{chi-3d-mlo}
\end{equation}
On assuming $\chi(z) = z^{\frac{l+1}{2}}(1-z)^{d}\phi(z)$, Eq. (\ref{chi-3d-mlo}) reduces to 
\begin{equation}
z(1 -z)\phi''(z) + \left[l+\frac{3}{2} - \left(l+2+2d\right)\right]\phi'(z) -\left(d + \frac{l + \tilde{\mu} + 1}{2}\right)
\left(d + \frac{l - \tilde{\mu} + 1}{2}\right) \phi(z) = 0,  
\label{chi-3d1-mlo}
\end{equation}
where $\tilde{\mu} = \sqrt{\mu^2 + 4\bar{\alpha\gamma}+(\bar{\alpha} - \bar{\gamma})^2}$. This is of the form of a hypergeometric equation and admits the solution at $x =0$ for (\ref{chi-3d1-mlo}) as
\begin{eqnarray}
\phi(z) &=& C_1\;{}_2{F}_{1}\left(d + \frac{l + \tilde{\mu} + 1}{2},d + \frac{l - \tilde{\mu} + 1}{2}; l+\frac{3}{2}; z\right) \nonumber \\
& &+ C_2\;z^{-l-\frac{1}{2}} {}_2{F}_{1}\left(d + \frac{-l + \tilde{\mu}}{2},d - \frac{l + \tilde{\mu}}{2};\frac{1}{2}-l, z\right), 
\; |z|<1.   
\label{phi3d-mlo}
\end{eqnarray} 
It reduces to polynomial form when $d + \frac{l + \tilde{\mu} + 1}{2} = -n_r$, 
\begin{equation}
\phi(z) = C_1\;{}_2{F}_{1}\left(-n_r,n_r - \tilde{\mu}; l+\frac{3}{2}; z\right) + C_2 z^{-l-\frac{1}{2}} {}_2{F}_{1}(-n_r-l-\frac{1}{2}, n_r-l-\tilde{\mu}+1;\frac{1}{2}-l, z), \; |z|<1,   
\label{phi3d1-mlo}
\end{equation} 
where $C_1$ and $C_2$ are integration constants, which sets  the energy eigenvalues (\ref{energy-mlo-3d1}) as  
\begin{eqnarray}
\hspace{-0.5cm} E_{n_r,l} = \left(2 n_r + l + \frac{3}{2}\right)\hbar\omega_0 \sqrt{1+\frac{1}{\mu^2}\left(\frac{9}{4}-2\eta_1\right)} - \left[\left(2 n_r + l+\frac{3}{2}\right)^2  -\bar{\alpha}-\bar{\gamma}-\frac{9}{4}\right]\frac{\hbar^2\;\lambda}{2}, \nonumber \\ n_r = 0, 1, 2, ....  
\end{eqnarray}

The eigenfunctions are obtained as 
\begin{equation}
\Psi_{n_r,l,m}(r,\theta, \phi) = {\cal N}_{n_r,l}\;\frac{1}{\sqrt{\lambda}}\;\frac{(\lambda\;r^2)^{\frac{l}{2}}}{(1 + \lambda r^2)^{d}}\;\phi\left(\frac{\lambda r^2}{\lambda r^2 + 1}\right)\;Y_{l,m}(\theta, \phi),\quad n_r = 0, 1, 2, ...,
\end{equation}
where $\phi(z)$ are hypergeometric polynomials obtained in Eq. (\ref{phi3d1-mlo}) and $Y_{l,m}(\theta, \phi)$ are spherical harmonics with angular momentum value $l$ (a positive integer) and azimuthal quantum number $m$ ($|m| \le l$) and ${\cal N}_{n_r,l}$ is the normalization constant. 

\subsubsection*{(b) Negative values of $\lambda$}
When $\lambda < 0$, the particle is confined within the region $\left(0, \frac{1}{\sqrt{|\lambda|}}\right)$. The quantum behaviour of the system (\ref{r3dim-mlo}) are analyzed in the following two regions, 
\begin{eqnarray}
\mbox{Region I}&:& 0 \leq r < \frac{1}{\sqrt{|\lambda|}}, \\
\mbox{Region II}&:&  |r| >\frac{1}{\sqrt{|\lambda|}}. 
\end{eqnarray}

Eq. (\ref{r3dim-mlo}) is solved for the value of $\lambda < 0$ as in the previous case and admits the energy eigenvalues, 
\begin{eqnarray}
\hspace{-0.5cm} E_{n_r,l} = \left(2 n_r + l + \frac{3}{2}\right)\hbar\omega_0 \sqrt{1+\frac{1}{\mu^2}\left(\frac{9}{4}-2\eta_1\right)} + \left[\left(2 n_r + l+\frac{3}{2}\right)^2  -\bar{\alpha}-\bar{\gamma}-\frac{9}{4}\right]\frac{\hbar^2\;|\lambda|}{2}, \nonumber \\
 \hspace{8cm} n_r =0, 1, 2, 3, ..., N, \label{r3dim-neg-mlo} 
\end{eqnarray}
where $N$ is an integer, and the eigenfunctions are obtained as 
\begin{equation}
\hspace{-0.5cm}\qquad \psi_n= \left\{\begin{array}{cc}
        {\cal N}_{n_r,l}\;\frac{1}{\sqrt{|\lambda|}}\;(-|\lambda|\;r^2)^{\frac{l}{2}}(1 -|\lambda| r^2)^{\frac{l+\tilde{\mu} + 1}{2}}\;\tilde{\phi}\left(|\lambda| r^2\right)\;Y_{l,m}(\theta, \phi),  \qquad 0 \leq r < \frac{1}{\sqrt{|\lambda|}},\\
    \hspace{-4cm} 0, \hspace{3cm} |r|>\lambda^{-\frac{1}{2}},\,\, n_r =0, 1, 2, 3, ..., N, 
              \end{array}\right.\label{sol-3d-mlo}
\end{equation}

where $\tilde{\phi(z)}$ are hypergeometric polynomials evaluated as 
\begin{equation}
\tilde{\phi}(z) = C_1\;{}_2{F}_{1}\left(-n_r,n_r +l+ \tilde{\mu}+3/2; l+\frac{3}{2}; z\right) + C_2 z^{-l-\frac{1}{2}} {}_2{F}_{1}(-n_r-l-\frac{1}{2}, n_r+\tilde{\mu}+1; \frac{1}{2}-l, z), \; |z|<1.   
\label{tphi3d1-mlo}
\end{equation}

The radial part given by Eq. (212) also has a continuous spectrum of states when $d$ is purely imaginary for both positive and negative  values of $\lambda$.
%\subsubsubsection{\bf Region II - $|r| >\frac{1}{\sqrt{|\lambda|}}$}

\subsection{{\label{qhigg3d} Quantum solvability of the three dimensional Higgs oscillator}}
To understand the quantum behaviour of the Higgs oscillator as described by the Lagrangian (\ref{schwingerq-higg}), we express the components of conjugate momenta in polar coordinates $(p_r, p_{\theta}, p_{\phi})$,   
\begin{equation}
p_r = \frac{\dot{r}}{(1 + k r^2)^2}, \quad p_{\theta} = \frac{r^2}{(1 + k r^2)}\dot{\theta}, \quad {\text and}\quad  p_{\theta} = \frac{r^2\;\sin^2{\theta}}{(1 + k r^2)}\dot{\phi}. 
\label{mom}
\end{equation}
We can derive the corresponding classical Hamiltonian in polar coordinates from the Lagrangian (\ref{schwingerq-higg}),  
\begin{eqnarray}
H = \frac{1}{2}\left[(1 + k r^2)^2 p_r^2 +\frac{(1 + k r^2)}{r^2}\left(p^2_{\theta} + \frac{p^2_{\phi}}{\sin^2{\theta}}\right) + 
\omega^2_0 r^2 \right]. \label{hamD3}
\end{eqnarray}
 This allows us to separate the radial and angular parts of the Hamiltonian, which simplifies the analysis of the system.   

Again by using the same transformation, ${\displaystyle \Psi(r, \theta, \phi) = \frac{\chi(r)}{r} Y(\theta, \phi)}$ and the procedure described in the previous section, the generalized Schr\"{o}dinger equation corresponding to the Hermitian ordered form (\ref{geham}) of the Hamiltonian (\ref{hamD3})  can be separated into 
\begin{eqnarray}
\hspace{-0.5cm}\;\;\mbox{Angular part:}&\;\;\;\;\;\;\;L^2 Y(\theta, \phi) = \hbar^2\;l(l+1)\; Y(\theta, \phi),& \label{a3dim}\\
\hspace{-0.5cm}\;\;\mbox{Radial part:}&\;\;{\displaystyle \chi''+\frac{4\;k r}{(1+k\;r^2)}\chi'+ \left[\frac{A}{(1 + k r^2)} + \frac{B}{(1+kr^2)^2} - \frac{l(l+1)}{r^2}\right]\chi = 0,} &
\label{r3dim}
\end{eqnarray}
where $Y_{l, m}(\theta, \phi) = \sqrt{\frac{(l - m)!}{(l + m)!}}\; e^{i m \phi}\; P^m_l(\cos{\theta})$ are spherical harmonics. As before, $P^m_l$ is the associated Legendre polynomial, $l$ is the angular momentum quantum 
number and $m$ is the magnetic quantum number. For every value of $l$, $m = -l, -(l-1), ... 0, ... l-1, l$, that is $m$ takes $2 l + 1$ values. Here 
\begin{eqnarray}
A &=& 2\eta_1 - k\;\mu^2 + k l(l+1),\\
B &=& -4\eta_2 + k\;\mu^2 + \frac{2 E}{\hbar^2},\label{energyd1}\\
\mu &=& \frac{\omega_0}{\hbar\;k}.\label{muh}
\end{eqnarray}

As we have done in the one dimensional case, we solve the radial part (\ref{r3dim}) for both $k > 0$ and $k<0$.
 
\subsubsection*{(a)\; Positive values of $k$}
When $k > 0$, $r$ spreads over from $0$ to $\infty$. On using a set of transformations on (\ref{a3dim}) and (\ref{r3dim}), we can solve the Schr\"{o}dinger equation (\ref{geham_heo3d}). The eigenfunctions are obtained in terms of the associated Jacobi polynomials as \cite{ruby2021classical}
\begin{equation}
\hspace{-0.5cm}\Psi_{n_r,l,m}(r,\theta, \phi) = {\cal N}_{n_r,l}\;\frac{1}{\sqrt{k}}\;\frac{(k\;r^2)^{\frac{l}{2}}}{(1 + k r^2)^{\frac{l}{2}+\frac{\mu}{2}+\frac{5}{4}}}\;P^{(\mu,\; l+1/2)}_{n_r}\left(\frac{k r^2 - 1}{k r^2 + 1}\right)\;Y_{l,m}(\theta, \phi),  
\end{equation}
where $P^{(a, b)}_{\nu}$ are the associated Jacobi polynomials and $Y_{l,m}(\theta, \phi)$ are spherical harmonics with angular momentum value $l$ and azimuthal quantum number $m$ and the normalization constant is obtained as
\begin{eqnarray}
{\cal N}_{n_r,l} &=& \left(\frac{2 k \sqrt{k}\;n_r! (2 n_r+\mu +l + 3/2) \Gamma(n_r+l+\mu+3/2)}{\Gamma(n_r+\mu+ 1) \Gamma(n_r+l+3/2)}\right)^{\frac{1}{2}}. 
\end{eqnarray}

The energy eigenvalues are obtained as 
\begin{equation}
E_{n_r,l} = \left(2 n_r + l + \frac{3}{2}\right)\hbar\omega_0 \sqrt{1+\frac{1}{\mu^2}\left(\frac{9}{4}-2\eta_1\right)}+ \left[\left(2 n_r + l+\frac{3}{2}\right)^2  + 2\bar{\alpha}+2\bar{\gamma}-\frac{1}{4}\right]\frac{\hbar^2\;k}{2}. 
\end{equation}

\subsubsection*{(b) Negative values of $k$}

When $k < 0$, the radius is confined within the region $\left(0, \frac{1}{\sqrt{|k|}}\right)$. The quantum behaviour of the system (\ref{r3dim}) are analyzed in the following two regions, 
\begin{eqnarray}
\mbox{Region I}&:& 0 \leq r < \frac{1}{\sqrt{|k|}}, \\
\mbox{Region II}&:&  |r| >\frac{1}{\sqrt{|k|}}. 
\end{eqnarray}

The Schr\"{o}dinger equation (\ref{geham_heo3d}) in the Region I results in the energy eigenvalues,   
\begin{equation}
E_{n_r,l} = \left(2 n_r + l + \frac{3}{2}\right)\hbar\omega_0 \sqrt{1+\frac{1}{\mu^2}\left(\frac{9}{4}-2 \eta_1\right)}- \left[\left(2 n_r + l+\frac{3}{2}\right)^2  + 2\bar{\alpha}+2\bar{\gamma}-\frac{1}{4}\right]\frac{\hbar^2\;|k|}{2}. 
\end{equation}
and the eigenfunctions, 
\begin{eqnarray}
\hspace{-0.5cm} \quad \Psi_{n,l}(r) &=& {\cal N}_{n,l}\;\frac{(|k|\;r^2)^{\frac{l+1}{2}}}{(1 - |k| r^2)^{n_r+\frac{l}{2}-\frac{\mu}{2}+\frac{5}{4}}}\;\left[C_1\;{}_2{F}_{1}\left(-n_r,n_r+\mu; l+\frac{3}{2}; |k| r^2\right) \right. \nonumber \\
\hspace{-0.5cm} & & \qquad\left.+ C_2 (|k| r^2)^{-l-\frac{1}{2}} {}_2{F}_{1}(-n_r-l-\tilde{\mu}, -n_r-l;\frac{1}{2}-l, |k| r^2)\right]Y_{l,m}(\theta, \phi). 
\label{chizk-}
\end{eqnarray}
The eigenstates are bounded if $n_r+\frac{l}{2}+\frac{5}{4} > \frac{\mu}{2}$ and $n_r =0, 1, 2, ...N$. 

Similarly for the region $II$, we can obtain the solution $\Psi_{n,l}(r)$ for  $|r|>\frac{1}{\sqrt{|k|}}$ as
\begin{eqnarray}
\hspace{-0.5cm}\Psi_{n,l}(r) &=& (|k|\;r^2)^{\frac{l+1}{2}}}{\left[C_1 (1-|k|r^2)^{-\frac{l}{2}-\frac{5}{4}+\frac{\tilde{\mu}}{2}}{}_2F_1\left(d+\frac{l}{2}-\frac{\tilde{\mu}}{2}+\frac{5}{4}, -d-\frac{l}{2}+\frac{\tilde{\mu}}{2}+\frac{1}{4}; 1-\tilde{\mu}; \frac{1}{1-|k| r^2}\right) \right. \nonumber \\
\hspace{-0.5cm} & &\left. + C_2 (1-|k|r^2)^{-\frac{l}{2}-\frac{5}{4}-\frac{\tilde{\mu}}{2}}{}_2F_1\left(d+\frac{l}{2}+\frac{\tilde{\mu}}{2}+\frac{5}{4}, -d-\frac{l}{2}+\frac{\tilde{\mu}}{2}+\frac{1}{4}; 1+\tilde{\mu}; \frac{1}{1-|k| r^2}\right) \right],\nonumber  \\ 
\hspace{-0.5cm} & &\times Y_{l,m}(\theta, \phi) \label{scatk-}
\end{eqnarray}
where $Y_{l,m}(\theta, \phi)$ are spherical harmonics and ${}_2{F}_{1}(a,b; c, z)$ are hypergeometric functions. The states (\ref{scatk-}) are continuum states. 
 
\subsection{Quantum solvability of $k$-dependent nonlinear oscillator}
The classical Hamiltonian corresponding to the Lagrangian (\ref{schwingerq-nlor}) in polar coordinates can be expressed as 
\begin{eqnarray}
H = \frac{1}{2}\left[(1 + k r^2)^2 p_r^2 +\frac{(1 + k r^2)}{r^2}\left(p^2_{\theta} + \frac{p^2_{\phi}}{\sin^2{\theta}}\right) + 
\frac{\omega^2_0 r^2}{(1 + k r^2)^2} \right]. \label{ham3dr-nl}
\end{eqnarray}

The generalized Schr\"{o}dinger equation associated with the non-hermitian ordered form of the Hamiltonian (\ref{ham3dr-nl}), which was used in the one-dimensional case, has the following form,  
\begin{eqnarray} 
\hspace{-0.5cm}\Phi''+ \left(\frac{2}{r}-(1+ \bar{\alpha} - \bar{\gamma})\frac{m'}{m}\right)\Phi'+ \left[\frac{(\bar{\gamma}-\bar{\alpha}-1)}{r}\frac{m'}{m}+\bar{\gamma} \frac{m''}{m} - \left(\overline{\alpha\gamma}+2\bar{\gamma} \right)\frac{m'^2}{m^2}\right. & & \nonumber \\
\hspace{-0.5cm} & &\hspace{-6cm}\left.-\frac{\sqrt{m}\;\hat{L}^2}{r^2}+ \frac{2\;m}{\hbar^2}\left(E - V(r)\right)\right]\Phi = 0. \label{geham_non3d}
\end{eqnarray}

Equation (\ref{geham_non3d}) can be separated by defining ${\displaystyle \Phi(r, \theta, \phi) = \frac{1}{r}\;\chi(r) Y(\theta, \phi)}$, with a spherically symmetric factor ${\displaystyle \frac{\chi(r)}{r}}$ and ${\displaystyle m(r) = \frac{1}{(1 + k r^2)^2}}$ and a generalized angular function $Y(\theta, \phi)$,  
\begin{eqnarray}
\hspace{-0.5cm}\mbox{Angular part:}&\;\;\;\;\;L^2 Y(\theta, \phi) = \hbar^2\;l(l+1)\; Y(\theta, \phi),& \label{a3dim-nlo}\\
\hspace{-0.5cm}\mbox{Radial part:}&{\displaystyle \chi''+\frac{4\;(\bar{\alpha}-\bar{\gamma}+1)\;k r}{(1+k\;r^2)}\chi'+ 
\left[\frac{4\sigma_1 k-k l(l+1)}{(1 + k r^2)} + \frac{\frac{2 E}{\hbar^2} + 4 \sigma_2\;k}{(1+kr^2)^2}  - \frac{\mu^2\;k r^2}{(1 + k r^2)^4} \right.}&\nonumber \\
\hspace{-0.5cm}&{\left. \displaystyle \hspace{8cm} - \frac{l(l+1)}{r^2}\right]\chi = 0,}& \nonumber \\ \label{r3dim-nlo1}
\end{eqnarray}
where $Y_{l, m}(\theta, \phi) = \sqrt{\frac{(l - m)!}{(l + m)!}}\; e^{i m \phi}\; P^m_l(\cos{\theta})$ are spherical harmonics, and $l,m$ are quantum numbers. 

By implementing the Bethe-Ansatz method, a quasi-exact treatment, on the equations (\ref{r3dim-nlo1}) and solving (\ref{a3dim-nlo}), one can obtain energy eigenvalues as
\begin{eqnarray}
\hspace{-0.5cm} \;E_{n,l} =\left(2 n + 2s +\frac{1}{2}\right) \hbar \omega_0 +\left(-\mu \sum^{n}_{i=1}z_i + \bar{\gamma} -( \bar{\gamma} - \bar{\alpha})(\bar{\gamma}-\bar{\alpha}-1)+\sigma_1\right)2\hbar^2 k,\label{enr-nlo}
\end{eqnarray}
with 
\begin{eqnarray}
\hspace{-0.5cm} \;\sigma_1 &=& \mu \sum^{n}_{i} z^2_i + (2- \mu)\sum^{n}_{i = 1} z_i-2n^2 -2s-\frac{l(l+1)}{4}+(\bar{\gamma}-\bar{\alpha})\left(\bar{\gamma}-\bar{\alpha}-\frac{3}{2}\right),
\label{sigman1_r}
\end{eqnarray}
 and the eigenfunctions for the system (\ref{geham_non3d}) are given by
\begin{eqnarray}
\hspace{-0.5cm}\;\Phi_{n,l}(r,\theta, \phi) = {\cal N}_{n,l}\;e^{\frac{-\omega_0 r^2}{2\;\hbar(1 + k r^2)}}(k r^2)^{\frac{l}{2}}(1 + k r^2)^{n + \bar{\gamma}-\bar{\alpha}} \Pi^{n}_{i=0}\left(\frac{k\;r^2}{(1 + k r^2)} - z_i\right),\label{phirnl}
\end{eqnarray}
where ${\cal N}_{n,l}$ is the normalization constant and the roots can be found out by using the relation, 
\begin{eqnarray}
\sum^n_{j\neq i} \frac{2}{z_i - z_j}= \frac{\mu z_i^2 + (2n-\mu) z_i + \left(l + \frac{3}{2}\right)}{z_i(z_i - 1)}. 
\label{bethe-ansatz-nlo2}
\end{eqnarray}
The energy eigenvalues (\ref{enr-nlo}) are also of quadratic nature similar to the Higgs oscillator and Mathews-Lakshmanan oscillator. While solving the system (\ref{schwingerq-nlor}) quantum mechanically the ordering parameters play important role since $\sigma_1$ and $\sigma_2$ (vide (\ref{sigma1}) and (\ref{sigma2})) are also being related with $n$. This is possible when the ordering parameters are treated as arbitrary. To obtain the eigenfunctions for different values of $n$, we have to find out the roots $z_i, \; i = 0, 1, 2, ...n$,\ by solving the equation in $z_i$ of degree $n+1$. Hence it is possible only to obtain a few states of the system (\ref{ham3dr-nl}) explicitly and it is quasi exactly solvable. 

\section{Li\'enard type-II nonlinear oscillator}

The nonlinear oscillators with linear damping are categorized as Li\'enard type-II nonlinear oscillators (\ref{lie2}). The Modified Emden equation, a mathematical model describing the behavior of certain
astrophysical systems, is an example for such Li\'enard type-II nonlinear oscillator \cite{chandrasekhar1957introduction}.  Numerous  works are devoted to study such nonlinear systems in the literature. However, finding a time-independent Hamiltonian description for these systems has been a challenging task. Ref.\cite{chandrasekar2005unusual} presents a new approach that successfully provides the time-independent Hamiltonian for further exploration of the classical as well as quantum behaviors of these nonlinear systems. 

\subsection{Classical Solvability}
The one dimensional modified Emden equation,  
\begin{equation}
\qquad \ddot{x} + k x \dot{x} + \frac{k^2}{9} x^{3} + \omega^2 x = 0, 
\label{mee}
\end{equation}
where overdot denotes differentiation with respect to $t$ and $k$ and $\omega^2$ are real  
parameters, is shown to possess a Lagrangian $L$ given by
\begin{eqnarray}
\hspace{-0.5cm} \qquad \qquad \; \; L =  \frac{27 \omega^6}{2 k^2}\left(\frac{1}{k\dot{x} + \frac{k^2}{3} x^2 + 3\omega^2}\right) + \frac{3\omega^2}{2 k} \dot{x} - \frac{9 \omega^4}{2 k^2},
\end{eqnarray}
and the Hamiltonian, 
\begin{eqnarray}
\hspace{-0.5cm} \;\;\; H(x, p) = \frac{9 \omega^4}{2 k^2}\left[2 - \frac{2 k }{3 \omega^2} p - 2 \left(1 - \frac{2 k}{3 \omega^2} p\right)^{\frac{1}{2}}+ \frac{k^2 x^2}{9\omega^2}\left(1 - \frac{2 k}{3 \omega^2} p\right)\right], \; -\infty < p \le \frac{3\omega^2}{2 k}, \nonumber \\ 
\label{ham1}
\end{eqnarray}
where the conjugate momentum is 
\begin{eqnarray}
\hspace{-0.5cm}\qquad \qquad \; \; p = \frac{\partial L}{\partial \dot{x}} = - \left( \frac{27 \omega^6}{2 k(k \dot{x}+\frac{k^2}{3} x^2 + 3 \omega^2)^2}\right) + \frac{3\omega^2}{2 k}.  
\label{mom}
\end{eqnarray}
The above Hamiltonian is of non-standard type, that is the coordinates and potentials
are mixed so that the Hamiltonian cannot be written as just the sum of the kinetic
and potential energy terms alone, including velocity dependent terms. 

\begin{figure}[t!]
\vspace{1cm}
\centering
\hspace{-1cm}\includegraphics[width=0.45\linewidth]{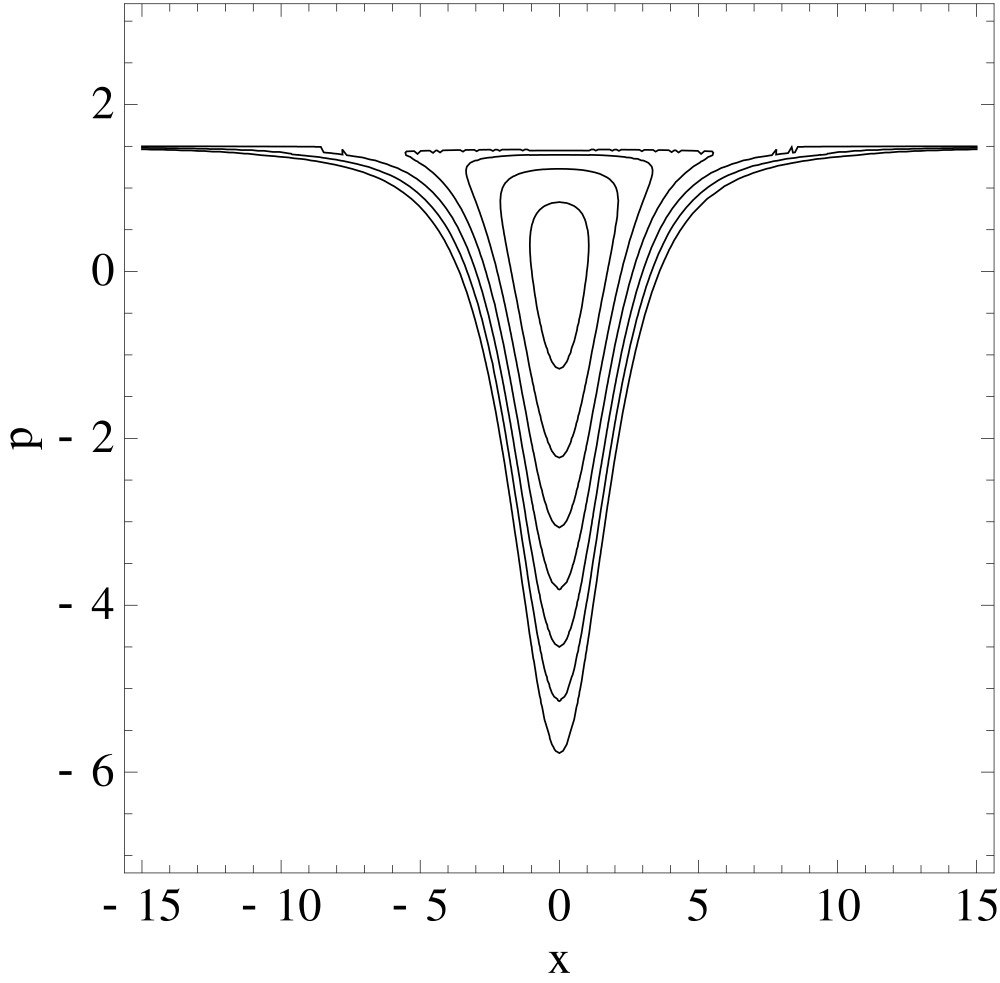}
\vspace{-0.1cm}
\caption{The phase trajectories of the Hamiltonian system (\ref{ham1}) with 
$\omega = k = 1$ for various values of $E = H$. }
\label{fig1-lie2}
\end{figure}

The phase space structure of the system (\ref{ham1}) is shown schematically in Figure \ref{fig1-lie2}.

The nonlinear oscillator (251) admits the general periodic solution which can be expressed as follows,
\begin{eqnarray}
\hspace{-0.5cm} \qquad \qquad \;\;\; x(t) = \frac{A \sin(\omega t + \delta)}{1 - \frac{k A}{3 \omega} \cos(\omega t + \delta)}, 
\quad \qquad 0 \le A < \frac{3 \omega}{k},
\label{solu1}
\end{eqnarray}
where $A,\; \delta$ are arbitrary constants. Note that for $0 \le A < \frac{3 \omega}{k}$
(while $-\infty < x < \infty$), the system (\ref{mee}) admits isochronous oscillations of frequency $\omega$,
which is the same as that of the linear harmonic oscillator.  For $A \ge \frac{3 \omega}{k}$, the 
solution is singular but periodic, while the corresponding momentum $p$ is 
bounded and periodic. 

\subsection{Quantum Solvability}
After performing a detailed analysis we find that the nonstandard classical Hamiltonian given in (\ref{ham1}) can also be equivalently considered in the form, \cite{ruby2012exact}, 
\begin{eqnarray}
 H(x, p) = \frac{x^2}{2\;m(p)} + U(p), \qquad -\infty < p \le \frac{3\omega^2}{2 k},
\label{pdms}
\end{eqnarray}
where
\begin{equation}
\;\;\;m(p) = \frac{1}{\omega^2\;\left(1 - \frac{2 k }{3 \omega^2} p\right)}
\;\;\;\mbox{and}\;\;\;U(p) = \frac{9\omega^4}{2 k^2} \left(\sqrt{1 - \frac{2 k}{3 \omega^2} p}-1\right)^2.
\label{mass}
\end{equation}

The observation of equivalence of the above Hamiltonian with position dependent mass Hamiltonian suggests to use different forms of ordered form of the Hamiltonian,\cite{von1983position,bastard1992monographies,gonul2002exact,kocc2003systematic}. We used von Roos ordered form to quantize of the Hamiltonian in momentum space with ${\displaystyle \hat{x} = i \hbar \frac{\partial}{\partial p}}$ and solved the corresponding time independent one dimensional Schr\"{o}dinger equation \cite{von1983position}. The procedure gives rise to unbounded solutions. Hence, we  proposed the non-Hermitian ordered form of the Hamiltonian (\ref{ham1}) associated with von Roos ordering in the form, 
\begin{eqnarray}
\hspace{-0.5cm} \; \tilde{H} &=& \frac{1}{\sqrt{m}} H \sqrt{m} = \frac{1}{2} \left[m^{-3/4}(\hat{p})\hat{x}m^{-1/2}(\hat{p})\hat{x}m^{1/4}(\hat{p})\right] + U(\hat{p}), \; \label{t_ham}\\
\hspace{-0.5cm} \;           &=& -\frac{\hbar^2}{2} \omega^2 \left(1 - \frac{2 k}{3 \omega^2} p \right) \left[ \frac{d^2}{d p^2} + \frac{k^2}{ 12 \omega^4} \frac{1}{\left(1 - \frac{2 k}{3 \omega^2} p \right)^2}\right] + \frac{9 \omega^4}{2 k^2} \left(\sqrt{\left(1 - \frac{2 k}{3 \omega^2} p \right)}  - 1\right)^2. \qquad  
\label{nham}
\end{eqnarray}
This Hamiltonian (\ref{nham}) is invariant under $\mathcal{PT}$-symmetry \cite{bender1998real, mostafazadeh2004physical, moiseyev2011non} though it is non-symmetric and non-Hermitian. 
The Schr\"{o}dinger equation corresponding to the Hamiltonian, $\tilde{H}(x, p)$  is 
\begin{eqnarray}
\hspace{-0.5cm} \; -\frac{\hbar^2 \omega^2 }{2}\left(1 - \frac{2 k}{3 \omega^2} p \right) \Phi'' - \frac{\hbar^2 k^2  }{ 24 \omega^2\left(1 - \frac{2 k}{3 \omega^2} p\right)} \Phi+ \frac{9 \omega^4}{2 k^2} \left(
\sqrt{1 - \frac{2 k}{3 \omega^2}p} - 1\right)^2 \Phi  = E \Phi, \;\;\left('= \frac{d}{d p}\right). \qquad 
\label{seh}
\end{eqnarray}
The eigenvalue problem associated with Eq. (\ref{seh}) can now be solved using the standard procedure. The resultant 
bound state solution turns out to be 
\begin{eqnarray}
\hspace{-0.5cm} \quad \Phi_n(p) = \left\{\begin{array}{cc} \tilde{N}_n \left(1 - \frac{2k}{3\omega^2}p\right)^{1/4} \exp{\left(-\frac{9\omega^{3}}{2\;\hbar\;k^2}\left(1 - \frac{2k}{3\omega^2}p - 2\sqrt{1 - \frac{2k}{3\omega^2}p}\right)\right)}\nonumber \\
\hspace{3cm} \times H_{n}\left[\frac{3\omega^{3/2}}{\sqrt{\hbar}\;k}\left(\sqrt{1-\frac{2k}{3\omega^2}p} - 1\right)\right], 
-\infty < p \le \frac{3 \omega^2}{2 k}, \label{sol2a} \\
                                  \hspace{-7cm}0, \quad p \ge \frac{3 \omega^2}{2 k}, 
                                   \end{array}\right. \nonumber \\
																	\label{sol2}
\end{eqnarray}
and the corresponding energy eigenvalues continue to be 
\begin{eqnarray}
\hspace{-0.5cm} \hspace{4cm}  E_{n} &=& \left(n+\frac{1}{2}\right)\;\hbar\; \omega, \qquad \quad n = 0, 1, 2, ....
\label{sol1}
\end{eqnarray}
It can be observed that the solution represented by equation \eqref{sol2} is continuous, single-valued, and bounded throughout the entire region $-\infty < p < \infty$. Furthermore, it satisfies the boundary conditions $\Phi(\infty) = \Phi({3 \omega^2}/{2 k}) = \Phi(-\infty) = 0$. The eigenfunctions $\Phi_n(p)$ are also bounded and continuous within the interval $-\infty < p \le \frac{3 \omega^2}{2 k}$ and are zero outside  this range \cite{ruby2012exact}. 

As a result, the above eigen functions  are normalizable and the associated normalization
constant $\tilde{N}_n$ is
\begin{eqnarray}
\tilde{N}_n = \left(\frac{e^{-\left(\frac{9 \omega^{3}}{k^2 \hbar}\right)}}{\sqrt{\hbar \omega}(2^{n-1} \sqrt{\pi} n! \left(1 + \frac{9 \omega^{3}}{k^2 \hbar}\right) + g(a))} \right)^{1/2}, 
\label{norm2}
\end{eqnarray}
where 
\begin{eqnarray}
g(a) =  \int^{\frac{3\omega^2}{2k}}_{0} \Phi^{*}_{n}(p)\; \Phi_{n}(p) dp. 
\end{eqnarray}

\subsubsection*{B. 1. \bf Broken $\mathcal{PT}$-Symmetry}
When the canonically conjugate momentum $p > \frac{3 \omega^2}{2 k}$,  the quantum Hamiltonian (\ref{nham}) is {\cal PT}-symmetric. The Schr\"{o}dinger equation corresponding to the potential (\ref{nham}) for the von Roos ordering admits  the solution, 
\begin{eqnarray}
\hspace{-0.5cm}\quad \Phi_n(p) = \left\{\begin{array}{cc} \tilde{N}_n \left(\frac{2k}{3\omega^2}p-1\right)^{1/4} \exp{\left(-\frac{9\omega^{3}}{2\;\hbar\;k^2}\left(\frac{2k}{3\omega^2}p -1 + i \; 2\sqrt{1 - \frac{2k}{3\omega^2}p}\right)\right)}\nonumber \\
\hspace{3cm} \times H_{n}\left[\frac{3\omega^{3/2}}{\sqrt{\hbar}\;k}\left(\sqrt{1-\frac{2k}{3\omega^2}p} + i\right)\right], 
-\infty < p \le \frac{3 \omega^2}{2 k}, \label{sol2a} \\
                                  \hspace{-7cm}0, \quad p \ge \frac{3 \omega^2}{2 k}, 
                                   \end{array}\right. \label{sol2-}
\end{eqnarray}
and the energy eigenvalues are obtained as 
\begin{eqnarray}
\hspace{-0.5cm} \hspace{4cm}  E_{n} &=& -\left(n+\frac{1}{2}\right)\;\hbar\; \omega, \qquad \quad n = 0, 1, 2, ....
\label{sol1}
\end{eqnarray}
Hence, the eigenfunctions cease to exhibit $\mathcal{PT}$-symmetry, even though the Hamiltonian possesses $\mathcal{PT}$-symmetry. Consequently, this leads to a negative energy spectrum that is unbounded below \cite{ruby2012exact}.

The quantization of the system (\ref{pdms}) has been accomplished through the application of supersymmetric quantum mechanics \cite{abdellaoui2018quantization}, employing a procedure that ensures the preservation of the Noether symmetry by the Hamiltonian  \cite{gubbiotti2014noether}. Very recently the modified Emden dynamical equation has been generalized to describe position-dependent mass (PDM) classical particles using a general nonlocal point transformation which has been further studied for its classical solvability \cite{mustafa2023dynamics}.

\section{Conclusion}
In this paper, we analyzed different types of Li\'{e}nard type nonlinear oscillators that have both linear and nonlinear damping terms. These oscillators exhibit a wide range of behaviors in both classical and quantum regimes. We first focused on one-dimensional Li\'{e}nard type I and type II oscillators and their associated Euler-Lagrange equations. The classical studies showed that the Li\'{e}nard type-I oscillators have localized solutions, isochronous and non-isochronous oscillations. Then we considered their quantum versions. In particular, Mathews-Lakshmanan and Higgs oscillators are quantum mechanically exactly solvable for arbitrary choice of ordering parameters. The ordering parameters play a crucial role in their solutions. On the other hand, some of the nonlinear oscillators have non-isochronous classical solutions expressed in terms of elliptic functions. In the quantum region, they are quasi-exactly solvable. 

We also explored the classical dynamics of the three-dimensional generalizations of these classical systems and their quantum equivalents. We solved the isotonic generalization of non-isochronous nonlinear oscillators both classically and quantum mechanically. This advancement in studies helps us to understand the behaviour of these systems more systematically.

The modified Emden equation is categorized as Li\'{e}nard type-II, and it exhibits isochronous oscillations at the classical level. This property makes it a valuable tool for studying the underlying nonlinear dynamics. We also examined its quantum equivalent, which has the same linear energy spectrum as a linear harmonic oscillator. We have also shown that it is a typical $\mathcal{PT}$-symmetric nonlinear oscillator, exhibiting fascinating behaviours at both the classical and quantum regimes.

In this article, we have essentially paid attention to discuss the integrability/solvability at the classical and quantum regimes associated with Li\'{e}nard type I and II systems only. However, as expected one can identify many more interesting facts associated with other Li\'{e}nard type systems mentioned in the introduction. Recent studies have shown that the Dunkl generalizations of the Higgs oscillator (\ref{ham-higgs}) and the $\delta$-type system (\ref{hamiltonian}) can be solved under the chosen ordering. The reflection term has a significant impact on the solution \cite{schulze2024closed}. It is possible that the Dunkl generalization of similar systems could result in exactly or quasi-exactly solvable systems belonging to the Li\'{e}nard system category. Furthermore, the modified Emden equation (Eq. \ref{mee}) has been generalized and a non-standard Hamiltonian proposed \cite{pradeep2009nonstandard}. The solvability of the corresponding quantum system can be determined based on its parameters. We hope to pursue these aspects in future.

\section*{Acknowledgment}
The research work of M. L. supported by a SERB National Science Professorship and he wishes to express his thanks to the Science and Engineering Research Board, Department of Science and Technology, Government of India for the award.


\section*{References}

\begin{thebibliography}{20}
\bibitem{lienard1928etude}
Li{\'e}nard A 1928 Revue G\'{e}n\'{e}rale de l'\'{E}lactricit\'{e} \textbf{23} 901--902

\bibitem{lins2006lienard}
Lins A, De Melo W, and Pugh C C 2006 \textit{Geometry and Topology: III Latin American School of Mathematics Proceedings of the School held at the Instituto de Matem{\'a}tica Pura e Aplicada CNPg Rio de Janeiro July 1976} (Springer) 335--357

\bibitem{jordan2007nonlinear}
Jordan D and Smith P 2007 \textit{Nonlinear Ordinary Differential Equations: An Introduction for Scientists and Engineers} (United States, OUP Oxford)

\bibitem{lakshmanan2013generating}
Lakshmanan M and Chandrasekar V K 2013 The European Physical Journal Special Topics \textbf{222} 665--688

\bibitem{mathews1974unique}
Mathews P M and Lakshmanan M 1974 Quarterly of Applied Mathematics \textbf{32}(2) 215--218

\bibitem{mathews1975quantum}
Mathews P M and Lakshmanan M 1975 Il Nuovo Cimento A (1965-1970) \textbf{26}(3) 299--316

\bibitem{higgs1979dynamical}
Higgs P W 1979 Journal of Physics A: Mathematical and General \textbf{12}(3) 309

\bibitem{venkatesan1997nonlinear}
Venkatesan A and Lakshmanan M 1997 Physical Review E \textbf{55}(5) 5134

\bibitem{ruby2021classical}
Chithiika Ruby V and Lakshmanan M 2021 Journal of Physics A: Mathematical and Theoretical \textbf{54} 385301


\bibitem{van1920theory}
van der Pol B 1920 Radio review \textbf{1} 701--710

\bibitem{van1934nonlinear}
van der Pol B 1934 Proceedings of the Institute of Radio Engineers \textbf{22}(9) 1051--1086

\bibitem{kovacic2011duffing}
Kovacic I and Brennan M J 2011 \textit{The Duffing Equation: Nonlinear Oscillators and Their Behaviour} (United Kingdom, John Wiley \& Sons)

\bibitem{mustafa2021isochronous}
Mustafa O 2021 \textit{The European Physical Journal Plus} \textbf{136} 1-17

\bibitem{pradeep2009dynamics}
Gladwin Pradeep R, Chandrasekar V K, Senthilvelan M, and Lakshmanan M 2009 Journal of Physics A: Mathematical and Theoretical \textbf{42} 135206

\bibitem{chandrasekar2012class}
Chandrasekar V K, Sheeba J H, Gladwin Pradeep R, Divyasree R S, and Lakshmanan M 2012 Physics Letters A \textbf{376} 2188--2194

\bibitem{tiwari2015isochronous}
Ajey K Tiwari, Devi Durga A, Gladwin Pradeep R, and Chandrasekar V K 2015 Pramana \textbf{85} 789--805

\bibitem{pradeep2009nonstandard}
Gladwin Pradeep R, Chandrasekar V K, Senthilvelan M, and Lakshmanan M 2009 Journal of Mathematical Physics \textbf{50}(5) 052901

\bibitem{musielak2008standard}
Musielak Z E 2008 Journal of Physics A: Mathematical and Theoretical \textbf{41} 055205

\bibitem{ruby2012exact}
Chithiika Ruby V, Senthilvelan M, and Lakshmanan M 2012 Journal of Physics A: Mathematical and Theoretical \textbf{45} 382002


\bibitem{bastard1992monographies}
Bastard G 1992 \textit{Monographies de PhysiqueWave Mechanics Applied To Semiconductor Heterostructures} (Les Editions de Physique, Les Ulis)

\bibitem{gonul2002exact}
G{\"o}n{\"u}l B, {\"O}zer O, G{\"o}n{\"u}L B, and {\"U}zg{\"u}n F 2002 \textit{Modern Physics Letters A} \textbf{17} 2453--2465

\bibitem{kocc2003systematic}
Ko{\c{c}} R and Koca M 2003 \textit{Journal of Physics A: Mathematical and General} \textbf{36} 8105

\bibitem{gora1969theory}
Gora T and Williams F 1969 \textit{Physical Review} \textbf{177} 1179

\bibitem{chithiika2015removal}
Chithiika Ruby V, Chandrasekar V K, Senthilvelan M, and Lakshmanan M 2015 \textit{Journal of Mathematical Physics} \textbf{56} 012103

\bibitem{karthiga2017quantum}
Karthiga S, Chithiika Ruby V, Senthilvelan M, and Lakshmanan M 2017 Journal of Mathematical Physics \textbf{58}(10) 102110

\bibitem{tiwari2013classification}
Ajey K Tiwari, Pandey S N, Senthilvelan M, and Lakshmanan M 2013 Journal of Mathematical Physics \textbf{54}(5) 053506

\bibitem{delbourgo1969infinities}
Delbourgo R, Salam A, and Strathdee J 1969 Physical Review \textbf{187} 1999

\bibitem{carinena2004non}
Cari{\~n}ena J F, Ranada M F, Santander M, and Senthilvelan M 2004 Nonlinearity \textbf{17} 1941

\bibitem{carinena2007quantum}
Cari{\~n}ena J F, Ranada M F, and Santander M 2007 Annals of Physics \textbf{322} 434--459

\bibitem{lakshmanan1975quantum}
Lakshmanan M and Eswaran K 1975 \textit{Journal of Physics A: Mathematical and General} \textbf{8} 1658

\bibitem{ranada2002harmonic}
Ra{\~n}ada M F and Santander M 2002 Journal of Mathematical Physics \textbf{43} 431--451

\bibitem{quesne2015generalized}
Quesne C 2015 Journal of Mathematical Physics \textbf{56} 012903

\bibitem{quesne2018deformed}
Quesne C 2018 Journal of Mathematical Physics \textbf{59} 042104

\bibitem{bonatsos1994deformed}
Bonatsos D, Daskaloyannis C, and Kokkotas K 1994 \textit{Physical Review A} \textbf{50} 3700

\bibitem{ballesteros2006universal}
Ballesteros {\'A} and Herranz F J 2006 \textit{Journal of Physics A: Mathematical and Theoretical} \textbf{40} F51

\bibitem{ballesteros2009maximal}
Ballesteros {\'A} and Herranz F J 2009 \textit{Journal of Physics A: Mathematical and Theoretical} \textbf{42} 245203

\bibitem{ballesteros2013anisotropic}
Ballesteros {\'A}, Herranz F J, and Musso F 2013 \textit{Nonlinearity} \textbf{26} 971

\bibitem{schulze2013position}
Schulze-Halberg A, Garc{\'\i}a-Ravelo J, Pacheco-Garc{\'\i}a C, and Gil J J P 2013 \textit{Annals of Physics} \textbf{333} 323--334

\bibitem{ranada2014quantum}
Ra{\~n}ada M F 2014 Journal of Mathematical Physics \textbf{55}(8) 082108

\bibitem{trabelsi2013classification}
Trabelsi A, Madouri F, Merdaci A, and Almatar A 2013 arXiv preprint arXiv:1302.3963

\bibitem{dunkl1991integral}
Dunkl C F 1991 Canadian Journal of Mathematics \textbf{43} 1213--1227

\bibitem{genest2013dunkl}
Genest V X, Ismail M E H, Vinet L, and Zhedanov A 2013 Journal of Physics A: Mathematical and Theoretical \textbf{46} 145201

\bibitem{genest2014dunkl}
Genest V X, Ismail M E H, Vinet L, and Zhedanov A 2014 Communications in Mathematical Physics \textbf{329} 999--1029

\bibitem{quesne2023rationally}
Quesne C 2023 Journal of Physics A: Mathematical and Theoretical \textbf{56} 265203

\bibitem{schulze2022bound}
Schulze-Halberg A 2022 Modern Physics Letters A \textbf{37} 2250178

\bibitem{ruby2021quantum}
Chithiika Ruby V and Lakshmanan M 2021 Journal of Physics Communications \textbf{5} 065007

\bibitem{arscott1995heun}
Arscott F M 1995 \textit{Heun's Differential Equations} (Clarendon Press)

\bibitem{midya2009generalized}
Midya B and Roy B 2009 \textit{Journal of Physics A: Mathematical and Theoretical} \textbf{42} 285301

\bibitem{schulze2013rational}
Schulze-Halberg A and Roy B 2013 \textit{Journal of Mathematical Physics} \textbf{54} 122104

\bibitem{zhang2012exact}
Zhang Y-Z 2012 \textit{Journal of Physics A: Mathematical and Theoretical} \textbf{45} 065206

\bibitem{quesne2017families}
Quesne C 2017 \textit{Journal of Mathematical Physics} \textbf{58} 052104

\bibitem{ruby2022quantum}
Chithiika Ruby V, Chandrasekar V K, and Lakshmanan M 2022 \textit{Journal of Physics Communications} \textbf{6} 085006

\bibitem{chandrasekhar1957introduction}
Chandrasekhar S 1957 \textit{An Introduction to the Study of Stellar Structure} (Courier Corporation)

\bibitem{chandrasekar2005unusual}
Chandrasekar V K, Senthilvelan M, and Lakshmanan M 2005 \textit{Physical Review E} \textbf{72} 066203

\bibitem{von1983position}
von Roos O 1983 \textit{Physical Review B} \textbf{27}(12) 7547

\bibitem{bender1998real}
Bender C M and Boettcher S 1998 \textit{Physical Review Letters} \textbf{80} 5243

\bibitem{mostafazadeh2004physical}
Mostafazadeh A and Batal A 2004 \textit{Journal of Physics A: Mathematical and General} \textbf{37} 11645

\bibitem{moiseyev2011non}
Moiseyev N 2011 \textit{Non-Hermitian Quantum Mechanics} ( Cambridge University Press)

\bibitem{abdellaoui2018quantization}
Abdellaoui A and Benamira F 2018 \textit{Physica Scripta} \textbf{94} 015201

\bibitem{gubbiotti2014noether}
Gubbiotti G and Nucci M C 2014 \textit{Journal of Nonlinear Mathematical Physics} \textbf{21} 248--264

\bibitem{mustafa2023dynamics}
Mustafa O 2023 \textit{Physica Scripta} \textbf{98} 125211

\bibitem{schulze2024closed}
Schulze-Halberg A 2024 \textit{International Journal of Modern Physics A} \textbf{39} 2450013


\end{thebibliography}
\end{document}